\documentclass[%
 reprint,
nofootinbib,
 amsmath,amssymb,
 aps,
]{revtex4-1}

\usepackage{graphicx}
\usepackage{dcolumn}
\usepackage{bm}

\usepackage{dsfont} 

\usepackage{tikz}
\usetikzlibrary{positioning}

\begin{document}


\title{States of Low Energy in bouncing inflationary scenarios in Loop Quantum Cosmology}

\author{Mercedes Mart\'in-Benito}
 \email{m.martin.benito@ucm.es}
\author{Rita B. Neves}%
 \email{rneves@ucm.es}
\affiliation{Departamento de F\'isica Te\'orica and IPARCOS, Universidad Complutense de Madrid, Parque de Ciencias 1, 28040 Madrid, Spain}

\author{Javier Olmedo}
 \email{javolmedo@ugr.es}
\affiliation{Departamento de F\'isica Te\'orica y del Cosmos, Universidad de Granada, Granada-18071, Spain}%

\date{\today}

\begin{abstract}
In generic Friedmann-Lema\^itre-Robertson-Walker spacetimes, States of Low Energy (SLEs) are defined to minimize the regularized energy density smeared along the time-like curve of an isotropic observer, which is specified via a smearing function. For every smearing function, SLEs are unique (up to a phase) and are shown to be exact Hadamard states. In this work, we investigate the viability of SLEs as the vacuum for cosmological perturbations in hybrid Loop Quantum Cosmology, motivated by the fact that SLEs have been shown to provide suitable vacua in models where a period of kinetic dominance precedes inflation. We find that there are two classes of smearing functions that can be seen as natural choices within this context, for which the corresponding SLEs and the resulting power spectra at the end of inflation are quite insensitive to the exact shape and support of the smearing function. Furthermore, a preliminary analysis of the tensor-to-scalar ratio and of the spectral index indicates as good an agreement with observations as that of standard cosmology.
\end{abstract}

\maketitle


\section{Introduction}

The cosmic microwave background (CMB) is an excellent test bench for the study of the physics of the large scale structure of our Universe, specially after the recent high-precision observations reported by Planck Collaboration \cite{Planck2018_parameters,Planck2018_inflation}. There is considerable consensus that the origin of this large scale structure is primordial and naturally explained by the paradigm of cosmological inflation. The key ingredients within this paradigm are a homogeneous and isotropic universe undergoing a finite and nearly exponential expansion at early times, together with quantum primordial perturbations in a suitable vacuum state (typically the Bunch-Davies vacuum) at the onset of this exponential expansion. Then, quantum fluctuations of primordial perturbations at the end of inflation (codified in their power spectra) induce a distribution of temperature and polarization anisotropies in the CMB with properties compatible with current observations. Despite the success of this paradigm, which combines simplicity and predictability, inflationary scenarios in general relativity (GR) usually ignore the pre-inflationary dynamics, being the main reason that soon before the onset of inflation one encounters the classical big-bang singularity, where most of the models break down. 

This is a good motivation to consider scenarios with a well-defined pre-inflationary dynamics, free of singularities.  Independently of the concrete pre-inflatonary dynamics considered, cosmological perturbations will in general reach the onset of inflation in an excited quantum state with respect to the Bunch-Davies vacuum, affecting the power-spectra at the end of the exponential expansion and, therefore, potentially leaving imprints in the CMB. Although this is true in general, the predictability of these scenarios clashes with the lack of unique criteria for the choice of vacuum state. Indeed, while during an exponential expansion the natural vacuum state is the Bunch-Davies vacuum, selected by the symmetries of the spacetime, under other evolutions, symmetries do not serve to select a preferred vacuum state. In other words, in general cosmological spacetimes there is no unique notion of particle.

This question has received some attention. Actually, there are several criteria proposing candidates of vacuum state in cosmological settings based on natural requirements and applicable in quite general situations. The most popular prescription \cite{parker1969}, based on a WKB approximation, yields the so-called adiabatic states. Other proposals have explored a Hamiltonian diagonalization \cite{fulling1979,Fahn:2018,Elizaga2019}, minimization of the renormalized stress-energy tensor \cite{agullo2015,handley2016}, and minimization of the uncertainty relations \cite{danielsson2002,Ashtekar:2016}. Other choices minimizing physical quantities but smeared along time-like curves have also been considered. A recent application in kinetically dominated (bouncing) cosmologies consists of minimizing the oscillations of the power spectrum within the whole kinetic dominated era \cite{nonosc,menava,Navascues:2021}. In these lines, one of the most interesting proposals is that of the States of Low Energy (SLEs) defined in \cite{Olbermann2007}. This work is based on the result of \cite{Fewster2000}, where it is shown that, unlike the instantaneous energy density (at one spacetime point), the renormalized energy density, smeared along a time-like curve using a point-splitting procedure, is bounded from below as a function of the state. Thus, a state can be found that minimizes this quantity.
The work of \cite{Olbermann2007} adapts this result to Friedmann-Lema\^itre-Robertson-Walker (FLRW) spacetimes, considering a smearing function supported on the worldline of an isotropic observer. A procedure is developed by making use of a fiducial state, and finding through a Bogoliubov transformation the unique (up to a phase) state that minimizes each mode's contribution to the smeared energy density, and therefore dubbed as the State of Low Energy (SLE) associated to the smearing function. Remarkably, SLEs are shown to be of Hadamard type.\footnote{ Another interesting proposal for a vacuum state valid beyond cosmological spacetimes is that of \cite{afshordi2012}, which when restricted to these settings shows some similarities to those of \cite{Olbermann2007}. However the resulting state is not of Hadamard type \cite{Fewster_2012}.}

Evidently, there is a dependency of SLEs on the smearing function, which compromises the universality of the proposal and raises the question of what is the relation between these and the natural vacua of maximally symmetric spacetimes. It is desirable that any procedure aiming at defining a privileged vacuum state should somehow single out the natural vacuum in such spacetimes. In Minkowski, SLEs are trivially found to be identical to the natural Minkowski vacuum, regardless of the choice of the smearing function. In de Sitter, where  the preferred vacuum is given by the Bunch-Davies vacuum, the question is more involved. In \cite{DegnerPhD2013} it is shown that, in the massive field case in de Sitter, SLEs converge to the Bunch-Davies vacuum in an appropriate limit of the support of the smearing function. Proposals adopting a similar philosophy \cite{afshordi2012,nonosc} also identify those well-known vacua. 

A recent investigation \cite{Niedermaier2020} has explored several properties of SLEs. In particular, it has determined that these states admit ultraviolet and infrared expansions. When applying the construction to primordial perturbations, these results allow for an analytical development of the asymptotic behaviors of their power-spectra at the end of inflation, from which it is found that these agree with observations in models where a period of kinetic dominance precedes inflation. 

Since this is precisely the case in some bouncing inflationary scenarios, such as Loop Quantum Cosmology (LQC), in this work we will explore the consequences of SLEs on these quantum cosmology settings. LQC applies the quantization techniques of Loop Quantum Gravity (LQG) to cosmological models \cite{LQCreview_Bojowald2005,LQCreview_Ashtekar2011,LQCreview_Banerjee2012,LQCreview_Agullo2016}. It has been successfully applied to FLRW spacetimes \cite{APS_PRL,APS_extended,Bentivegna2008,Kaminski2009,Pawlowski2012} through its improved dynamics prescription \cite{APS_extended}, leading to a resolution of the big-bang singularity in terms of a quantum bounce that connects a contracting epoch of the Universe with an expanding one. Two main strategies have been followed to introduce perturbations in this formalism: the hybrid approach \cite{hybrid_FernandezMendez2012,hybrid_FernandezMendez2013,hybrid_FernandezMendez2014,hybrid_Gomar2014,hybrid_Gomar2015,hyb-ms,hybrid_Martinez2016,hyb-vs-dress,hyb-obs} and the dressed metric approach \cite{dressedmetric_Agullo2012,dressedmetric_Agullo_PRD_2013,dressedmetric_Agullo_CQG_2013,instvac}. In both approaches, corrected semiclassical equations that govern the evolution of cosmological perturbations are obtained. Several studies on the choice of vacuum state for perturbations within LQC and comparison with data have been carried out \cite{instvac,nonosc,menava,Navascues:2021,Ashtekar:2016}.

Here, we will focus our study on the hybrid approach, though the procedure can be reproduced within the dressed metric approach as well. In this work we propose SLEs as candidates for such vacua in LQC. The motivation is 2-fold: they provide a minimization of the regularized energy density (of each mode), and they are exact Hadamard states, which guarantees that computations such as that of the expectation value of the renormalized stress-energy tensor will be well defined. Furthermore, within LQC, we will show that the predictions of power spectra are very insensitive to the choice of smearing function of the SLEs, provided it covers the Planckian regime. By computing the tensor-to-scalar ratio and the spectral index, we also show that this choice of vacuum state can provide at least as good an agreement with observations as standard cosmological models incorporating the same inflaton potential. Besides, for the sake of completeness, we include a proof showing that, in de Sitter, SLEs prescription selects the Bunch-Davies vacuum state also in the case of a massless scalar field. In addition, one can resort to the Hadamard property of SLEs to explore whether other proposals for vacuum state are of Hadamard type as well. This is the case for the non-oscillatory vacuum states considered in Refs. \cite{nonosc, menava, Navascues:2021}, as we have recently proven in Ref. \cite{mno-sles}.

This manuscript is organized as follows. In section \ref{sec:intro pert} we review the dynamics of cosmological perturbations in inflationary FLRW models and the computation of their primordial power spectra. Section \ref{sec:introSLE} summarizes the procedure introduced in \cite{Olbermann2007} to obtain SLEs in an arbitrary FLRW model. In section \ref{sec:SLEdeSitter} we apply the procedure to the massless field in de Sitter and show that it converges to the Bunch-Davies vacuum in an appropriate limit. Section \ref{sec:SLEinLQC} is devoted to the computation of SLEs in LQC. Finally, in section \ref{sec:conclusions} we conclude  summarizing our results and with some closing remarks.

Throughout we adopt Planck units $c=\hbar=G=1$ for numerical computations, though factors of $G$ are kept in expressions.

\section{Cosmological perturbations}\label{sec:intro pert}

We will start by reviewing the dynamics of cosmological perturbations in an inflationary scenario. Let us consider a spatially flat FLRW spacetime, with lapse $N$ and scale factor $a$, minimally coupled to the massive scalar field $\phi$, subject to the potential $V(\phi)$. Throughout this work, we will mainly use conformal time $\eta$, fixing the lapse $N = a$, 
but cosmological time $t$ with lapse $N = 1$ will also be adopted, specially in section \ref{sec:introSLE} when referring to the notation of Olbermann in \cite{Olbermann2007}. Cosmological perturbations are commonly described by scalar and tensor gauge invariant perturbations, typically denoted by $Q$ and ${\cal T}^I$ (where $I$ encodes the two possible polarizations of the tensor modes). Performing a redefinition of the fields $u = a Q$ and $\mu^I=a {\cal T}^I$, and expanding in Fourier modes
\begin{equation}
    u(\eta,\vec{x}) = \frac{1}{(2\pi)^{3/2}} \int d^3k\,u_{\vec{k}}(\eta)e^{i\vec{k}\vec{x}},
\end{equation}
and equivalently for $\mu^I$, we find that, generally, each mode obeys
\begin{align}
    u^{\prime\prime}_{\vec{k}}(\eta) + \left(k^2+s^{(s)}(\eta)\right) u_{\vec{k}}(\eta) &= 0,\label{eq:eom_uk}\\
    \left(\mu^I_k(\eta)\right)^{\prime\prime} + \left(k^2+s^{(t)}(\eta)\right) \mu^I_{\vec{k}}(\eta) &= 0,\label{eq:eom_muk}
\end{align}
where prime denotes derivative with respect to $\eta$, $k = |\vec{k}|$, and $s^{(s)}(\eta)$ and $s^{(t)}(\eta)$ are the time-dependent masses of scalar and tensor modes, respectively.

In the particular case of the classical FLRW model 
\begin{equation}
    s^{(s)}(\eta)=-\frac{z^{\prime\prime}}{z},\qquad s^{(t)}(\eta)=-\frac{a^{\prime\prime}}{a},
\end{equation}
where $z=a\dot{\phi}/H$, and $H=\dot a/a$ is the Hubble parameter.  The dot represents derivative with respect to cosmological time $t$, related to conformal time through $dt=ad\eta$. As will be discussed in section \ref{sec:SLEinLQC},  $s^{(s)}$ and $ s^{(t)}$  take more complicated forms in LQC.
However, these time-dependent massess are very similar both in the classical and the quantum theory for kinetically dominated universes, where $z^{\prime\prime}/z=a^{\prime\prime}/a$, and also during slow-roll inflation. Therefore, in the rest of this section, we will focus only on scalar modes, as the discussion will be analogous to tensor modes.

Furthermore, since the field $u(\eta,\vec{x})$ is real, the Fourier modes satisfy  $ u^*_{\vec{k}}(\eta)=u_{-\vec{k}}(\eta)$,  where the asterisk denotes complex conjugation. Besides, in the equation of motion \eqref{eq:eom_uk} only modes with the same  wave number $k$ are coupled. Therefore, in the following, we will identify the Fourier modes with $u_{k}(\eta)$ for all $\vec k$ with the same modulus $k$. Each of the modes are normalized with respect to the usual Klein-Gordon inner product, which is time independent. Namely, given two complex solutions, $u_k^{(1)}$ and $u_k^{(2)}$ of the equation of motion \eqref{eq:eom_uk}, one can easily verify that 
\begin{equation}
\langle u^{(1)}_{k},u^{(2)}_{k}\rangle=-i\left\{u^{(2)}_{k}\left[\left(u^{(1)}_{k}\right)^{\prime}\right]^{*}-\left(u_{k}^{(1)}\right)^{*} \left(u_{k}^{(2)}\right)^{\prime}\right\},    
\end{equation}
is time independent. Thanks to this inner product, and following the usual strategy for the Fock quantization of perturbations on these time-dependent settings, we construct the one-particle Hilbert space out of a basis $u_{k}$ of solutions with positive (unit) norm
\begin{equation}
u_{k}\left(u_{k}^{\prime}\right)^{*}-\left(u_{k}\right)^{*} u_{k}^{\prime}=i,
\end{equation}
and their complex conjugate $u_{k}^{*}$ (which span the negative norm sector of the theory). Besides, solutions associated with wave vectors $\vec k$ and $\vec k'$, respectively, and such that $\vec k\neq\vec k'$, are orthogonal.

Let us note that we can parametrize the choice of basis of solutions in terms of suitable (normalized) initial conditions. In general,  up to a phase, they can be parametrized as
\begin{equation}\label{eq:CkDk}
    u_k(0) = \frac{1}{\sqrt{2 D_k}}, \qquad u_k^{\prime}(0) = \sqrt{\frac{D_k}{2}}\left(C_k-i\right),
\end{equation}
where $D_k$ is a positive function of $k$, and $C_k$ any real function of $k$. 

Given any initial conditions above, the perturbations can be evolved mode by mode with the equation of motion, until the time $\eta_{\text{end}}$ when the relevant scales have all crossed out the horizon.  After that the power spectra of the comoving curvature perturbation ${\cal R}_k=u_{k}/z$ and tensor modes will remain frozen. They are given by
\begin{align}
    \mathcal{P}_{\mathcal{R}}(k) &= \frac{k^{3}}{2\pi^{2}}\frac{|u_{k}|^{2}}{z^{2}}\Big|_{\eta = \eta_{\rm end}},\label{eq:PS_R}\\
    \mathcal{P}_{\mathcal{T}}(k) &= \frac{32k^{3}}{\pi}\frac{|\mu_{k}^{I}|^{2}}{a^{2}}\Big|_{\eta = \eta_{\rm end}},\label{eq:PS_T}
\end{align}
respectively.

\section{States of Low Energy}\label{sec:introSLE}

In \cite{Olbermann2007}, SLEs are defined as the ones that minimize the energy density smeared along a time-like curve. The work of \cite{Fewster2000} showed that this quantity, unlike the energy density in one spacetime point, has a lower bound and thus can be used to construct a class of states that turn out to have suitable properties for a vacuum state \cite{Niedermaier2020}. An appealing feature is that these are exact Hadamard states. In this section we will summarize the procedure of \cite{Olbermann2007} to define these states for an arbitrary FLRW model (its application in LQC scenarios will be explained in Sec. \ref{sec:SLEinLQC}). We refer the reader to \cite{Niedermaier2020}, where this procedure is studied in more detail. 

SLEs are built from the result of \cite{Fewster2000}, which uses a point splitting procedure to show that the energy density smeared along a time-like curve has a lower bound when considering Hadamard states. This then yields a well-defined expression for differences in smeared energy density on the time-like curve, which in \cite{Olbermann2007} is particularized for an isotropic observer in a homogeneous state.

Let us follow the notation of \cite{Olbermann2007} and consider $T_k(t)$, the mode decomposition of the minimally coupled field (which can represent in particular scalar or tensor modes), with equation of motion
\begin{equation}\label{eq:eom_Tk}
    \ddot{T}_k + 3H(t) \dot{T}_k + \omega^2_k(t) T_k = 0,
\end{equation}
where we remind that the dot represents derivative with respect to cosmological time $t$. Particularizing for scalar or tensor modes of primordial perturbations,  the time-dependent function $\omega_k$ relates to the notation of the previous section as
\begin{equation}
    \omega^2_k(t) = \frac{k^2 + s(t)}{a^2(t)} + H^2(t) + \frac{\ddot{a}(t)}{a(t)},
\end{equation}
where  $s(t)$ is the corresponding time-dependent mass term  expressed in cosmological time.

The procedure consists of finding the state, or equivalently, the solution $T_k$ which has minimal energy density associated to the smearing function $f$. This is achieved by minimizing the contribution of each mode $k$ to the total smeared energy density. For this purpose, it is more convenient to work within the Hamiltonian framework. The  contribution of each mode to the smeared energy density will be given by
\begin{equation}\label{energy}
   E(T_k) = \frac{1}{2} \int dt\,f^2(t)\left(\frac{|\pi_{T_k}|^2}{a^6} +\omega_k^2 |T_k|^2\right),
\end{equation}
where $\pi_{T_k}$ is the conjugate momentum of $T_k$, which is related to the velocity by means of $\pi_{T_k} = a^3\dot T_k$ via Hamilton's equations. As investigated in \cite{Olbermann2007}, we can start by considering a fiducial solution $S_k$ to \eqref{eq:eom_Tk}. Then, a generic solution to the equation of motion can be written by a Bogoliubov transformation as
\begin{equation}\label{eq:S_k to T_k}
    T_k = \lambda(k) S_k + \mu(k) \bar{S}_k,
\end{equation}
where $\lambda(k),\mu(k) \in \mathds{C}$ are the Bogoliubov coefficients with $|\lambda(k)|^2-|\mu(k)|^2 = 1$. Noting that there is a freedom in the choice of the complex phase of $T_k$ (if $T_k$ is a solution, then so is $e^{i \delta(k)} T_k$, with $\delta(k) \in \mathds{R}$), we can choose $\mu(k) \in \mathds{R}^+$ without loss of generality. This leaves us with only the following freedom: $\mu(k)$ and the complex phase of $\lambda(k)$, which we denote as $\alpha(k)$. Writing $E(T_k)$ in terms of the reference solution $S_k$ yields
\begin{equation}\label{eq:W(T_k)}
    E(T_k) = (2 \mu^2(k)+1) c_1(k) + 2\mu(k) \text{Re}[\lambda(k) c_2(k)],
\end{equation}
where, for a given fiducial solution $S_k$, $c_1(k)$ and $c_2(k)$ are fixed as
\begin{align}
    c_1(k) &:= \frac{1}{2} \int dt\,f^2(t)\left(\frac{|\pi_{S_k}|^2}{a^6} +\omega_k^2 |S_k|^2\right),\label{eq:c1}\\
    c_2(k) &:= \frac{1}{2} \int dt\,f^2(t)\left(\frac{\pi_{S_k}^2}{a^6} +\omega_k^2 S_k^2\right),\label{eq:c2}
\end{align}
with $\pi_{S_k}=a^3\dot S_k$. Since $\mu_k>0$, it is straightforward to see that a minimum $E(T_k)$ will require $\alpha(k) = \pi-\text{Arg}[c_2(k)]$. Then, minimizing $E(T_k)$ with respect to $\mu(k)$ yields:
\begin{align}
    \mu(k) &= \sqrt{\frac{c_1(k)}{2\sqrt{c_1^2(k)-|c_2^2(k)|}}-\frac{1}{2}}\ ,\label{eq:mu}\\
    \lambda(k) &= -e^{-i\text{Arg}[c_2(k)]} \sqrt{\frac{c_1(k)}{2\sqrt{c_1^2(k)-|c_2^2(k)|}}+\frac{1}{2}}\ .\label{eq:lambda}
\end{align}

In summary, the SLE associated to a smearing function $f$ is given (up to a phase) by \eqref{eq:S_k to T_k},  starting from a fiducial state defined via  the modes $S_k$, with $\mu(k)$ and $\lambda(k)$ found by \eqref{eq:mu} and \eqref{eq:lambda}, respectively. 

In \cite{Olbermann2007} it is also shown that for ultrastatic models there exists a state (unique up to a phase) which minimizes the energy density for all test functions, being dubbed the state of \emph{minimal} energy. In non ultrastatic models, however, such a state does not exist, every SLE is associated to a smearing function. We will explore this dependency in more detail in the following sections. 

Before we move to the next section, it is worth highlighting several important properties of the SLEs discussed in Refs. \cite{Olbermann2007, Niedermaier2020}. Besides the fact the SLEs are Hadamard states, their construction is gauge invariant with respect to the choice of lapse function, namely, they are time reparametrization invariant. Furthermore, Ref. \cite{Niedermaier2020} devises an alternative construction of such states, showing explicitly the independence of the fiducial solution of the equation of motion of the perturbations used to construct the SLEs, here denoted by $S_k$. It is shown that SLEs only depend on the commutator function, which is state independent.  

\section{SLE in de Sitter}\label{sec:SLEdeSitter}

As we have just seen, SLEs are not unique inasmuch as they depend on the choice of the test function $f$. Then one might wonder whether there is any relation between these states and the natural vacuum present in maximally symmteric spacetimes, namely Minkowski and de Sitter. These spacetimes  admit a unique vacuum invariant under the isometries of the spacetime, the Minkowski vacuum and the Bunch-Davies vacuum respectively. It is therefore desirable that any prescription attempting to define the vacuum, when applied to these spacetimes, singles out such a state.

In Minkowski this questions is trivial.  Indeed, if we take as fiducial solution the Minkowski vacuum, it is obvious that the integrand in \eqref{eq:c2} is identically zero for any $f$, and thus the SLE is found to be the Minkowski vacuum (up to a phase). This is not the case in a de Sitter universe. Thus, the question arises whether the SLEs converge to the Bunch-Davies vacuum in some appropriate limit. Ref. \cite{DegnerPhD2013} proves that indeed this is the case for the massive field theory.

For the sake of completeness in this discussion, in this section we will extend the analysis to the massless case, which is also common in cosmology. Let us remind that the cosmological chart of de Sitter is defined by the space-time metric
\begin{align}
    ds^2=a^2(\eta)[-d\eta^2+d\vec{x}\,^2]\quad,\quad a(\eta)=-\frac{1}{H\eta},
\end{align}
where conformal time takes values $\eta\in(-\infty,0]$, and  the Hubble parameter $H$ is constant. 
In this cosmological chart of de Sitter, even though strictly speaking we cannot define a unique vacuum state invariant under the isometries of the spacetime, cosmologists typically consider as natural vacuum the one defined by 
\begin{align}\label{bdvacuum}
T^{BD}_k(\eta)=-\frac{H}{\sqrt{2k}}e^{-ik\eta}\left( \eta-\frac{i}{k}\right),
\end{align}
that we will keep calling Bunch-Davies state. Let us introduce a test function $f(\eta)$ defining the time-like curve of an isotropic observer in de Sitter,  with support in the interval $[\eta_0,\eta_f]$. Then, the smeared energy density of each mode in the vacuum \eqref{bdvacuum} measured by that observer is 
$E(T^{BD}_k)$ defined in \eqref{energy} with $\omega_k^2=k^2/a^2$. Using conformal time, it explicitly reads
\begin{align}\label{ebd}
  E(T^{BD}_k)=\frac{H^3}{2}\int_{\eta_0}^{\eta_f} d\eta f^2(\eta)|\eta| \left(k\eta^2+\frac{1}{2k} \right).
\end{align}

Let us now consider any other state, defined via $T^\omega_k$.
Such a state will be related to the Bunch-Davies state through a Bogoliubov transformation that, up to an irrelevant global phase, is given by
\begin{align}
T^{\omega}_k=e^{i\alpha(k)} \sqrt{1+\beta^2(k)} T^{BD}_k+\beta(k) \bar{T}^{BD}_k,
\end{align}
with  $\alpha(k)$ and $\beta(k)\neq 0$ real coefficients. 
Then, we can write
\begin{align}
    E(T^{\omega}_k)=E(T^{BD}_k)(1+2\beta^2) +2\beta\sqrt{1+\beta^2}\text{Re}[e^{i\alpha} D(T_k^{BD})],
\end{align}
with
\begin{equation}
   D(T_k) = \frac{1}{2} \int dt\,f^2(t)\left(\dot{T}_k^2 +\omega_k^2 T_k^2\right),
\end{equation}
which implies
\begin{align}
 D(T_k^{BD})=-\frac{H^3}{2}\int_{\eta_0}^{\eta_f} d\eta f^2(\eta)|\eta|e^{-2ik\eta} \left(i\eta+\frac{1}{2k} \right).
\end{align}

Let us now show that, no matter what this state is, it always verifies $E(T^{\omega}_k)\geq E(T^{BD}_k)$ in the limit $\eta_0\rightarrow-\infty$ and for any time-like curve of the observer measuring the state's energy. The  smallest value for $E(T^{\omega}_k)$ is attained by choosing $\alpha(k)$ such that $e^{i\alpha(k)} D(T_k^{BD})=-|D(T_k^{BD})|$ and $\beta(k)>0$. In that case, and defining $\delta(k)=\beta(k)/\sqrt{1+\beta(k)^2}\in(0,1)$, we have the following relation 
\begin{align}
   E(T^{\omega}_k)=\mathcal{R}_k(\eta_f,\eta_0)E(T^{BD}_k),
\end{align}
with
\begin{align}
   \mathcal{R}_k(\eta_f,\eta_0)=\left\{1+\frac{2\delta(k)}{1-\delta(k)^2}\left[\delta(k) -\frac{|D(T_k^{BD})|}{E(T^{BD}_k)}\right]\right\}
\end{align}

It is easy to see that ${|D(T_k^{BD})|}/{E(T^{BD}_k)}$ is bounded by 
\begin{align}\label{ratio}
 \frac{|D(T_k^{BD})|}{E(T^{BD}_k)}\leq \frac{\int_{\eta_0}^{\eta_f} d\eta f^2(\eta)\eta^2\sqrt{1+\frac{1}{(2k\eta)^2} }}{\int_{\eta_0}^{\eta_f} d\eta f^2(\eta)\eta^2\left(k|\eta|+\frac{1}{2k|\eta|} \right)}.
\end{align}
Then, in the limit $\eta_0\rightarrow-\infty$, the denominator in \eqref{ratio} grows faster than the numerator no matter the test function $f$ and we have
\begin{align}
   \lim_{\eta_0\to-\infty}\mathcal{R}_k(\eta_f,\eta_0)= \left[1+\frac{2\delta^2(k)}{1-\delta^2(k)}\right]>1\quad\forall \,k,\eta_f.
\end{align}

This implies
   \begin{align}
   E(T^{\omega}_k)> E(T^{BD}_k)\,,\quad \forall\,k
\end{align}
independently of the choice of both $f$ and the final point of its support, provided that the initial point $\eta_0$ tends to the distant past. 

This clarifies the relation between SLEs in the case of the massless field in the cosmological chart of de Sitter and the Bunch-Davies vacuum: they agree whenever the observer starts measuring the energy at the distant past, since then the Bunch-Davies vacuum has the smallest possible energy. This result actually agrees with the conclusion found also in the massive case \cite{DegnerPhD2013}.

\section{SLE{\lowercase{s}} in LQC}\label{sec:SLEinLQC}

The work of \cite{Niedermaier2020} has shown that SLEs are an appropriate choice of vacuum for cosmological perturbations in models where a period of kinetic dominance precedes inflation. This is precisely the typical case of some inflationary bouncing cosmologies, like LQC. Here, several investigations have proposed choices of initial vacua \cite{instvac,nonosc,Ashtekar:2016,menava} for scalar and tensor perturbations that provide power spectra compatible with observations, including power suppression at large angular scales (for an appropriate number of $e$-folds). However, to our knowledge, SLEs have not been explicitly analyzed in the context of LQC. Therefore, as a working example, we will study this prescription in LQC through its hybrid approach \cite{hyb-vs-dress,hyb-ms,hyb-obs}.

Here, the Fourier modes of gauge invariant scalar and tensor perturbations satisfy the equations of motion \eqref{eq:eom_uk} and \eqref{eq:eom_muk} respectively, with time-dependent mass terms that depend on the quantization approach adopted (see Ref. \cite{hyb-vs-dress}). In the case of the hybrid approach, they can be written in terms of the background variables $a$, $\rho$ (inflaton energy density), $P$ (inflaton pressure) and the inflaton potential $V(\phi)$ as follows:
\begin{align}
    s^{(t)} &= -\frac{4\pi G}{3} a^2 \left(\rho - 3P\right),\\
    s^{(s)} &=  s^{(t)} + \mathcal{U},
\end{align}
where
\begin{equation}\label{eq:U_MS}
    \mathcal{U} = a^2\left[V_{,\phi\phi}+48\pi G V(\phi)+ 6 \frac{a^{\prime}\phi^{\prime}}{a^3 \rho} V_{,\phi}-\frac{48\pi G}{\rho} V^2(\phi)\right].
\end{equation}

Generally, there is no analytical solution to \eqref{eq:eom_uk} and \eqref{eq:eom_muk} with such time-dependent mass terms. These can be solved numerically, given initial conditions $u_k(\eta_0)$, $u^{\prime}_k(\eta_0)$, for scalar modes, and $\mu^I_k(\eta_0)$, $[\mu^I_k(\eta_0)]^{\prime}$ for tensor modes, uniquely specifying a choice of vacua for perturbations.

To apply the procedure outlined in Sec. \ref{sec:introSLE}, we first obtain a fiducial solution $S_k$, by fixing initial conditions at the bounce, and then we integrate the dynamics numerically. We recall that the fiducial solution considered is irrelevant since it does not affect our final SLE. Therefore, for simplicity we fix $S_k$ at the bounce to be the 0th order adiabatic state,  defined choosing $D_k=k$ and $C_k=0$. We refer the reader to Appendix \ref{sec:adiabatic} for a brief reminder on adiabatic vacua.
Furthermore, for these computations, we have considered a quadratic inflaton potential: $V(\phi)= m^2 \phi^2/2$, with $m=1.2 \times 10^{-6}$ in Planck units. In the following we will adopt these units unless otherwise specified. We choose this value of the mass to get agreement with observations, inspired by previous studies in LQC \cite{AshtekarSloan_probinflation}. Finally, we also need to fix the free parameter of the background, namely $\phi_B$, the value of the inflaton field at the bounce. Since we do not intend to do here a rigorous Bayesian analysis, we choose $\phi_B$ such that the deviation of scale invariance in the power spectra only affects the largest scales, with respect to those permitted by observations, where the cosmic variance is not small. Some tuning is necessary, as too low values will result in too little inflation, pushing the oscillations of the power spectra to the range where they are excluded by observations, and too big values will lead to power spectra that show no deviation from the standard model in the scales of interest. In this spirit, we fix $\phi_B = 1.225$ in Planck units, though a range of values around it would still produce the desired qualitative behavior.

Then one has to fix the test function $f^2$. Let us focus initially on the choice of its support. In standard cosmology, it is common to fix the initial time for the vacuum of perturbations at the onset of inflation, since the classical theory breaks down closely before, at the big bang singularity. On the other hand, LQC offers a singularity-free geometry, where curvature never blows up but reaches a maximum Planckian magnitude. Hence, it seems natural to consider initial conditions at this high-curvature region. However, one can also select an initial time in the asymptotic past well before the bounce. These have mainly been the two strategies followed in the literature.

Noticeably, for the SLE construction, we do not need to fix an initial time. As long as we have a solution $S_k$ (for all $\eta$), the choice of which is irrelevant, we can construct the solution $T_k$ (for all $\eta$) that provides the minimal smeared energy density. However, the freedom mentioned above is now replaced in part by the choice of the test function. Initially, one might reckon that, since we are minimising the energy density as measured by an observer, it seems more natural to allow this observer to witness the whole evolution, by which we mean choosing $f^2$ with a wide enough support around the bounce (since it should be compactly supported). Notwithstanding, one could argue that this is a naive view, since LQC provides quite simplified scenarios. For instance, cosmological models derived from full LQG provide a quantum bounce that, instead of connecting two low curvature FLRW spacetimes, displays a collapsing branch that is indeed a de Sitter spacetime with a Planckian cosmological constant (see for instance Refs. \cite{Dapor:2017,Assanioussi:2018,Assanioussi:2019,quismondo:2019,Agullo:2018,Li:2019,Li:2020,Olmedo:2018}). It is therefore  reasonable to explore test functions with support that includes the contracting and expanding branch, as well as those corresponding to a (smooth but steep) window function from the bounce point to the future, so that only the expanding branch is considered.

In each of these two classes of test functions, we find that the resulting SLE does not depend on their support, as long as it is wide enough. In other words, the SLE converges very quickly with the support of $f^2$. Moreover, for the first class, where $f^2$ has support on a very wide period from before the bounce to the future, the resulting SLE does not depend on the shape of the test function either\footnote{ We computed several possibilities, with different test functions, for instance considering the bump function and combining different number of bumps one after the other, even playing around with their amplitudes, and our computations showed convergence to essentially the same SLE.}. Therefore, for the sake of simplicity, in this manuscript we will show the results for $f^2$ being a (smooth) window function. The computations were done in conformal time $\eta$, through the change of variables $dt = a d\eta$ in the integrals of \eqref{eq:c1} and \eqref{eq:c2}. For this reason, we define the window function in terms of conformal time directly by making use of the auxiliary function
\begin{equation}
    S(x) = \frac{1-\tanh\left[\cot(x)\right]}{2},
\end{equation}
such that $f^2$, supported in the interval $\eta \in \left[\eta_0,\eta_f\right]$, is defined as:
\begin{equation}
    f^2(\eta) = 
\begin{cases}
	S\left(\frac{\eta-\eta_0}{\delta}\pi\right) &\eta_0\leq \eta < \eta_0+\delta,\\
	1 & \eta_0+\delta\leq \eta \leq \eta_f-\delta,\\
	S\left(\frac{\eta_f-\eta}{\delta}\pi\right) &\eta_f-\delta < \eta \leq \eta_f,
\end{cases}
\end{equation}
where $\delta$ controls the ramping up, with small $\delta$ resulting in a steeper step. The first situation that we will consider, which we will name the whole evolution one, is studied by choosing this function with large enough support around the bounce. Concretely, we have determined that for a range around the bounce $\eta_f-\eta_0 = 64$ (Planck seconds) the resulting SLE has already converged. The second situation, which we will refer to as the expanding branch, is investigated by fixing $f^2$ to be this window function starting at the bounce and ending at the onset of inflation, with a very small value of $\delta$, so that the contribution from the dynamics close to the bounce is not dampened. Specifically, we have chosen $\delta \sim 0.06$ (Planck seconds). In this second situation we did not consider anymore different shapes of the test function as, provided that its support does not suppress the kinetically dominated regime, the results are not going to depend much on the particular shape, similarly to the behavior that we observed in the first situation.

Figure \ref{fig:CD} shows the value at the bounce of the functions $D_k$ and $C_k$ that characterize the SLEs for scalar modes obtained with these two strategies. Though it might seem surprising at first, the procedure gives SLEs with almost exactly the same initial conditions at the bounce for tensor modes, which is why we have omitted that plot. However, it is easy to realize that this has to be the case. Firstly, let us note that, close to the bounce, the time-dependent masses of the two types of perturbations are very similar, as their difference, given by \eqref{eq:U_MS}, is subdominant for a kinetically dominated bounce. As mentioned, the SLE is independent of the fiducial solution $S_k$, and so it is in particular independent of the initial conditions imposed to obtain $S_k$. In this case we are free to choose for $S_k$ the same initial conditions for scalar and tensor modes (as is indeed the case with the 0th order adiabatic ones we have adopted). Then it is easy to see that, given the same initial conditions and very close equations of motion at and around the bounce, the procedure should provide a state that maintains a great similarity between the value of the state at the bounce for scalar and tensor modes.
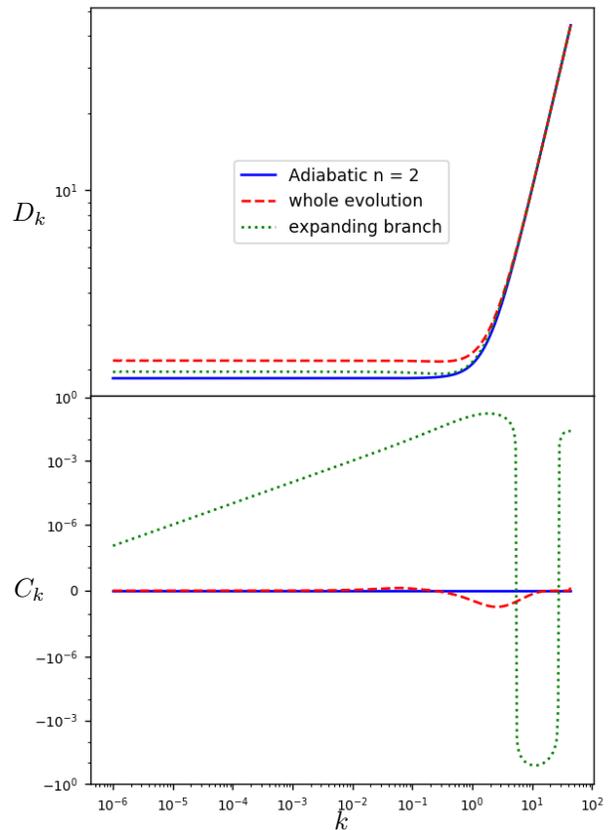
\begin{figure}
    \begin{tikzpicture}
    \tikzstyle{every node}=[font=\normalsize]
    \node (img)  {\includegraphics[width=0.45\textwidth]{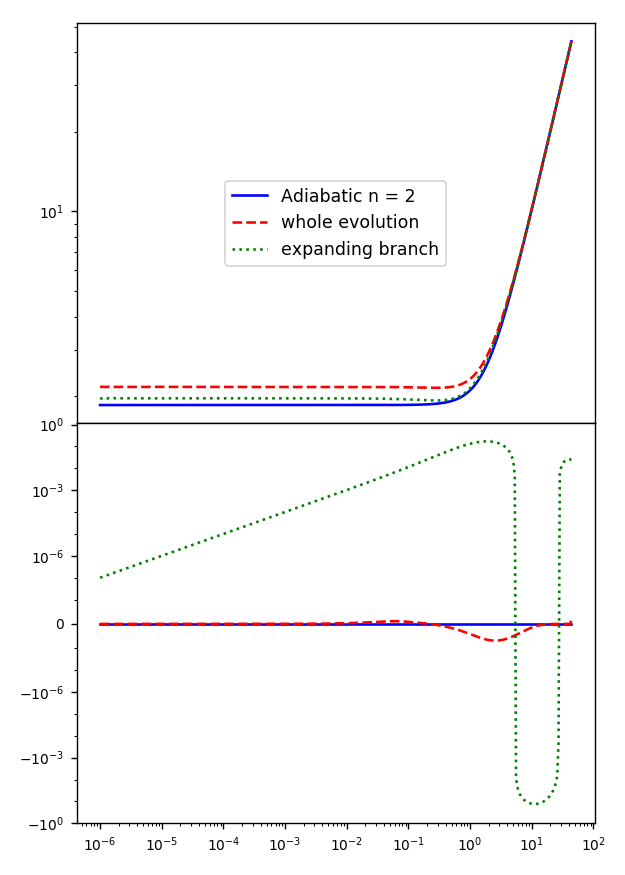}};
    \node[left=of img, node distance=0cm, anchor=center,yshift=2.6cm,xshift=1.3cm] {$D_k$};
    \node[left=of img, node distance=0cm, anchor=center,yshift=-2.4cm, xshift=1.3cm] {$C_k$};
    \node[below=of img, node distance=0cm, anchor=center,yshift=1.3cm, xshift=.3cm] {$k$};
 \end{tikzpicture}
    \caption{Initial conditions for scalar modes at the bounce corresponding to the SLEs obtained with window functions covering the whole evolution (dashed red lines) and only the expanding branch (dotted green lines) in terms of $D_k$ and $C_k$, as constructed in \eqref{eq:CkDk}. Second order adiabatic initial conditions computed through \eqref{eq:ad_n} are also shown for comparison (blue line). The scale of $k$ is in Planck units. All computations were performed for a quadratic potential with $m=1.2\times 10^{-6}$, and with $\phi_B = 1.225$. For tensor modes, the resulting SLE at the bounce shows no significant qualitative differences.}
    \label{fig:CD}
\end{figure}
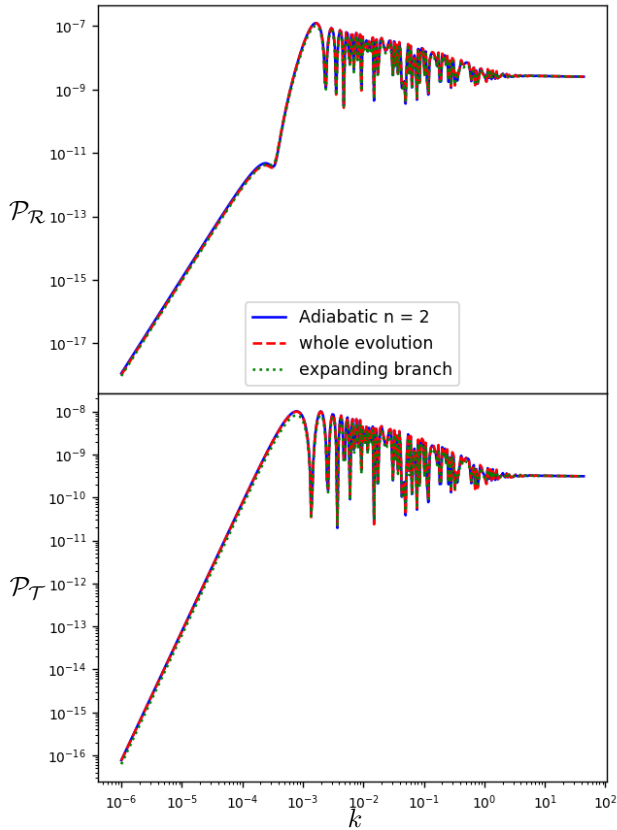
\begin{figure}
    \begin{tikzpicture}
    \tikzstyle{every node}=[font=\normalsize]
    \node (img)  {\includegraphics[width=0.45\textwidth]{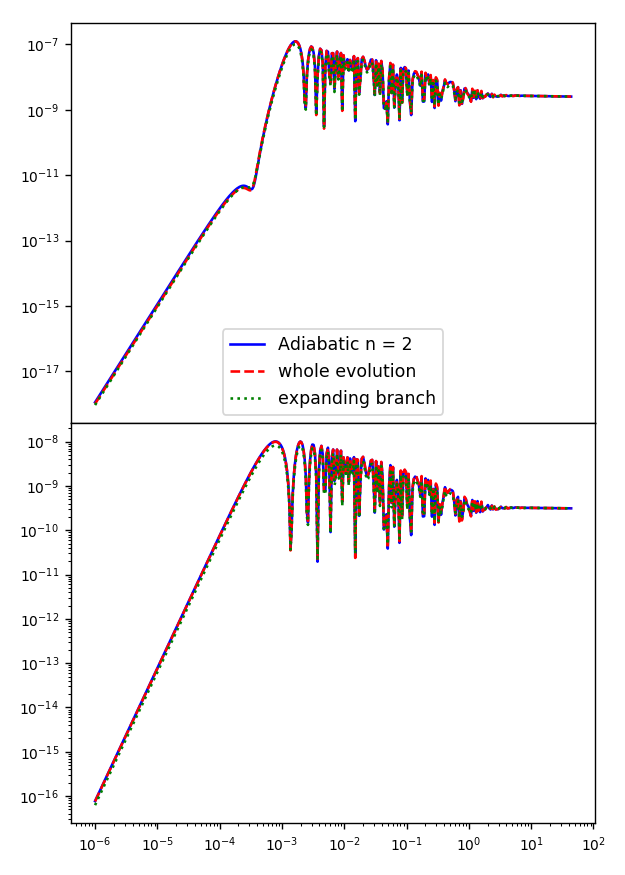}};
    \node[left=of img, node distance=0cm, anchor=center,yshift=2.6cm,xshift=1.1cm] {$\mathcal{P}_{\mathcal{R}}$};
    \node[left=of img, node distance=0cm, anchor=center,yshift=-2.4cm, xshift=1.1cm] {$\mathcal{P}_{\mathcal{\mathcal{T}}}$};
    \node[below=of img, node distance=0cm, anchor=center,yshift=1.3cm, xshift=.3cm] {$k$};
 \end{tikzpicture}
    \caption{Power spectra of the comoving curvature perturbation $\mathcal{P}_{\mathcal{R}}$ and tensor perturbation  $\mathcal{P}_{\mathcal{T}}$ corresponding to the SLEs obtained with window functions covering the whole evolution (dashed red lines) and only the expanding branch (dotted green lines). The ones for second order adiabatic initial conditions at the bounce are also shown for comparison (blue line). The scale of $k$ is in Planck units. All computations were performed for a quadratic potential with $m=1.2\times 10^{-6}$, and with $\phi_B = 1.225$.}
    \label{fig:PS}
\end{figure}
Regarding the comparison between the two choices of test function, Figure \ref{fig:CD} shows that $D_k$ has the same kind of behavior in both cases, though it tends to a different value in the infrared. On the other hand, $C_k$ reveals an entirely different behavior between the two scenarios, even though the magnitude in all cases is smaller than unity.

Remarkably, we have found that both scenarios yield extremely similar power spectra at the end of inflation, as is shown in Figure \ref{fig:PS}. It seems that the differences observed in initial conditions have no measurable impact in predictions of power spectra. Pairing this to the convergence of the SLE with respect to the form and support of the test function mentioned above, one may say that observational predictions seem to be manifestly insensitive to the choice of test function, within these natural scenarios in LQC.

Interestingly, we find that the power spectra match extremely well with the one obtained with second order adiabatic initial conditions at the bounce, as computed from \eqref{eq:ad_n}, which is also shown in Figure \ref{fig:PS} for comparison. However, it is important to stress that the SLEs are fundamentally different from adiabatic initial conditions as choices of vacuum. Firstly, SLEs minimize the smeared energy density. Secondly, they are exact Hadamard states, which is not the case for adiabatic states of finite order. Importantly, this allows for the computation of quantities such as the regularized stress-energy tensor. It is worth pointing out that, even if one is not interested in such computations and would therefore be more inclined to use simpler constructions for initial conditions such as adiabatic ones, it would not be necessary to go any further than second order so as to approximate the SLE.

\begin{figure}
    \begin{tikzpicture}
    \tikzstyle{every node}=[font=\normalsize]
    \node (img)  {\includegraphics[width=0.45\textwidth]{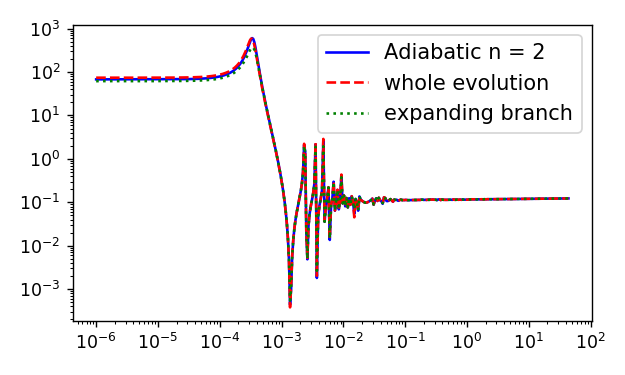}};
    \node[left=of img, node distance=0cm, anchor=center,yshift=0cm,xshift=1.2cm] {$r$};
    \node[below=of img, node distance=0cm, anchor=center,yshift=1.3cm, xshift=.3cm] {$k$};
 \end{tikzpicture}
    \caption{Tensor-to-scalar ratio corresponding to the SLEs obtained with window functions covering the whole evolution (dashed red line) and only the expanding branch (dotted green line). The one for second order adiabatic initial conditions at the bounce is also shown for comparison (blue line). The scale of $k$ is in Planck units. All computations were performed for a quadratic potential with $m=1.2\times 10^{-6}$, and with $\phi_B = 1.225$.}
    \label{fig:r}
\end{figure}

For the sake of completeness, in Figure \ref{fig:r} we show the tensor-to-scalar ratio ($r=\mathcal{P}_{\mathcal{T}}/\mathcal{P}_{\mathcal{R}}$) as a function of $k$. As we see, this observable is scale invariant even at scales where the scalar and tensor power spectra depart from scale invariance. This is in agreement with previous results reported in the literature \cite{hyb-obs,dressedmetric_Agullo_CQG_2013}. 

Finally, to roughly locate our results in the context of observations of the CMB, we have computed two quantities: $r_{0.002}$, the value of the tensor-to-scalar ratio at $k = 0.002\ \text{Mpc}^{-1}$, and the spectral index $n_s$. We note that this is only a first investigation with toy values for the model, and that these computations serve merely to show the potential of using SLEs as vacuum states for primordial perturbations.
Furthermore, we note that the scale of $k$ shown in the figures of this manuscript is in Planck units and corresponds to the usual choice of fixing the scale factor to be $1$ at the bounce in LQC. To relate these to observations of the CMB (namely to identify the scale corresponding to $k = 0.002\ \text{Mpc}^{-1}$ where the tensor-to-scalar ratio is to be computed), which fix the scale factor to be $1$ today, one needs first to identify the correspondence between $k$ of observations, which we will denote as $\tilde{k}$, with that of our own model, which we will keep calling $k$. This is accomplished by identifying the pivot scale $k^{\star}$ that corresponds to the one of observations $\tilde{k}^{\star}$. For greater detail on this matter in LQC see for example \cite{dressedmetric_Agullo_CQG_2013}. In this work, we will compare our results to the ones obtained by the Planck Collaboration \cite{Planck2018_inflation}. For the pivot scale of the Planck Collaboration of $\tilde{k}^{\star} = 0.05 \text{Mpc}^{-1}$, we find the corresponding one in our model in Planck units to be $k^{\star} = 43.9$ when $\phi_B = 1.225$. Then, it is easy to find that to $\tilde{k} = 0.002\ \text{Mpc}^{-1}$ corresponds $k \simeq 1.76$.
We find the tensor-to-scalar ratio at this scale for the test function supported along the whole evolution, $r_{0.002}^{\text{w}}$, and for it supported on the expanding branch only, $r_{0.002}^{\text{e}}$, as the ratio between the tensor and scalar power spectra at $k \simeq 1.76$:
\begin{equation}
    r_{0.002}^{\text{w}} \simeq 0.118, \qquad r_{0.002}^{\text{e}} \simeq 0.117.
\end{equation}
These values seem to be somewhat disfavored by the observations reported by the Planck collaboration. However they perfectly agree with the predictions for standard cosmology with the same quadratic potential for the scalar field, also shown in \cite{Planck2018_inflation}. In this sense, the disparity with observations likely stems from the choice of this potential and not from the choice of vacuum for the perturbations.

The computation of the spectral index is more straightforward. It is found by fitting the power spectrum of the comoving curvature perturbation with the function:
\begin{equation}
    \mathcal{P}_{\mathcal{R}} = A_S \left(\frac{k}{k^{\star}}\right)^{n_s-1},
\end{equation}
where $A_S$ is the value of $\mathcal{P}_{\mathcal{R}}$ at $k = k^{\star}$. Concretely, we have computed $n_s$ by fitting the power spectrum with this function in a range in the ultraviolet such that shifting this range to lower or higher values of $k$ made no difference in the final result of $n_s$ to three decimal points. This way we find, for the two possibilities of the test function:
\begin{equation}
    n_s^{\text{w}} \simeq n_s^{\text{e}} \simeq 0.969.
\end{equation}
This value agrees with mean values and error bars from observations from the Planck collaboration.

Finally, it is worth mentioning that we expect the SLEs to lead to different predictions than the standard Bunch-Davies vacuum in GR and previous proposals in LQC. For instance, in Ref. \cite{agullo2015}, the resulting vacuum state that yields a vanishing (renormalized) stress-energy tensor at the bounce is a 4th order adiabatic state that produces a stronger enhancement of power at small wave numbers compared to the SLEs. On the other hand, the vacua proposed in Refs. \cite{nonosc,Ashtekar:2016} show suppression of power at those scales, rather than enhancement, with respect to the Bunch-Davies power spectrum. We leave for future work a detailed comparison between the predictions coming from our proposal and those of previous ones.

\section{Conclusions and discussion} \label{sec:conclusions}

In \cite{Olbermann2007}, a class of states of a field minimally coupled to the geometry is identified in generic FLRW models as the ones that minimize the regularized energy density smeared along a time-like curve. These are named the SLEs associated to the smearing function, and are shown to be of Hadamard type. Subsequently, they have been shown to be viable candidates for vacuum states of cosmological perturbations. The fact that they are Hadamard states is a great advantage, as it guarantees that computations such as that of the  expectation value of the renormalized stress-energy tensor will be well defined. Furthermore, the construction of concrete Hadamard states or even the proof that a state is of Hadamard type are often rather complicated. Additionally, in \cite{Niedermaier2020} they have also been shown to provide a qualitative agreement between predictions of power spectra at the end of inflation and observations for infrared and ultraviolet scales, in models where a period of kinetic dominance precedes inflation.

In general cosmological scenarios, these states are not free of ambiguities. By definition, they minimize the smeared energy density with respect to a given test function. Hence, one might question their universality as natural vacua for perturbations. To this respect, we have pointed out that the analysis of SLEs in maximally symmetric spacetimes seems to indicate that they do indeed select the preferred vacuum when a clear notion of one already exists. In Minkowski, for example, the situation is trivial and the SLE is immediatly identified as the Minkowski vacuum. In de Sitter, in Ref. \cite{DegnerPhD2013} it was proven that SLEs select the Bunch-Davies vacuum in an appropriate limit of the test functions in the case of the massive theory. For completeness, we have also shown in section \ref{sec:SLEdeSitter} that this is the case in the massless theory as well. These results  suggest that the ambiguity in the test function might be surpassable at least if a natural choice of test function is identified.

When applied to cosmological perturbations in LQC, we have identified two natural choices for the test function: one supported along a wide enough window around the bounce, and one that includes the expanding branch only and is steep so that the bounce contribution is not dampened. In both cases we find that the resulting SLE does not depend on the support of the test function as long as it is wide enough. For the first choice of support, we further find that the SLE does not depend on the shape of the test function. Remarkably, we have found that the SLEs yield power spectra at the end of inflation that do not depend qualitatively on either choice as long as the test function has support on the high-curvature region of the geometry. Even quantitatively this dependence is very week. Then, the major disadvantage in selecting SLEs as vacuum states for primordial perturbations, namely that of the ambiguity in the test function, effectively disappears at least in the context of hybrid LQC  for these natural choices. We expect this would be a robust feature within other approaches for cosmological perturbation in LQC as well. Furthermore, a  preliminary analysis of the tensor-to-scalar ratio and the spectral index shows that these are at least as viable candidates for vacuum states of perturbations as any other previous proposals regarding their agreement with observations, with the additional advantage that they are of Hadamard type.

Finally, let us stress that the goal of this work is to present SLEs as suitable candidates for vacua of perturbations in LQC. This has been accomplished by demonstrating that there exists a naturally motivated class of test functions for which the results seem to be insensitive to their particular form and support. A preliminary analysis has also shown that in this case an agreement with observations is attainable. However, to determine this agreement rigorously, we plan to carry out a proper Bayesian analysis, studying not only the possible parameters of the test functions, but also the free parameters coming from the LQC framework, such as the value of the inflaton field at the bounce. In addition, these states can also be interesting in order to alleviate tensions of anomalies in the CMB, in the line of \cite{Agullo:2020,Agullo:2020b,Ashtekar:2020,Ashtekar:2021}.
We further intend to investigate the viability of SLEs as vacua of perturbations in other approaches of cosmological perturbations within LQC, such as the dressed metric approach \cite{dressedmetric_Agullo_CQG_2013}, or other cosmological models obtained from LQG \cite{Dapor:2017,quismondo:2019,Agullo:2018,Olmedo:2018}, or alternative  treatments for cosmological perturbations within LQG \cite{Han:2020,Schander4:2019}, and compare the results between different approaches.

\acknowledgments

We acknowledge I. Agullo and A. Ashtekar for stimulating discussions. This work is supported by the Spanish Government through the projects FIS2017-86497-C2-2-P and PID2019-105943GB-I00 (with FEDER contribution). R.~N. acknowledges financial support from Funda\c{c}\~ao para a Ci\^encia e a Tecnologia (FCT) through the research grant SFRH/BD/143525/2019. J. O. acknowledges the "Operative Program FEDER2014-2020 Junta de Andaluc\'ia-Consejer\'ia de Econom\'ia y Conocimiento" under project E-FQM-262-UGR18 by Universidad de Granada.

\appendix

\section{Adiabatic vacua}\label{sec:adiabatic}

Among the possible candidates for initial conditions, adiabatic vacua are one of the most popular choices for selecting a set of positive frequency solutions.\footnote{They were originally proposed as approximated solutions to the equations of motion although nowadays it is common to consider them for the selection of initial conditions for cosmological perturbations.} To define them, one first considers the ansatz:
\begin{equation}
    u_k(\eta) = \frac{1}{\sqrt{2W_k(\eta)}}e^{-i\int^{\eta}W_k(\bar{\eta}){\rm d}\bar{\eta}},
\end{equation}
which is plugged into the equations of motion of the perturbations, yielding:
\begin{equation}\label{eq:W_adiabatic}
    W_k^2 = k^2 + s(\eta) -\frac{1}{2}\frac{W^{\prime\prime}_k}{W_k}+\frac{3}{4}\left(\frac{W_k^{\prime}}{W_k}\right)^2,
\end{equation}
where we have used generically $s(\eta)$ to denote the time-dependent mass term.

A solution of order $n$ denoted by $W_k^{(n)}$ is defined as an approximated solution that converges to $W_k$ in the limit of large $k$, at least as $\mathcal{O}\left(k^{n-\frac{1}{2}}\right)$.

There is not a unique procedure to obtain the functions $W_k^{(n)}$. We will adopt the following one: the adiabatic solution of order $n+2$, namely $W_k^{(n+2)}$, is obtained by inserting in the right-hand side of \eqref{eq:W_adiabatic} the solution $W_k^{(n)}$:
\begin{equation}\label{eq:ad_n}
    \left(W_k^{(n+2)}\right)^2 = k^2 + s(\eta) -\frac{1}{2}\frac{W^{(n)\prime\prime}_k}{W^{(n)}_k}+\frac{3}{4}\left(\frac{W_k^{(n)\prime}}{W^{(n)}_k}\right)^2,
\end{equation}
starting with $W_k^{(0)} = k$. In this paper we will rather use adiabatic solutions as initial data that will be evolved with the exact equations of motion. We will therefore refer to the state defined by the set of solutions with initial data 
\begin{equation}
    D_k = W_k^{(n)}(\eta_0), \qquad C_k = -\frac{[W_k^{(n)}(\eta_0)]^{\prime}}{2 [W_k^{(n)}(\eta_0)]^2},
\end{equation}
for some initial time $\eta_0$ as an adiabatic vacuum of order $n$. We will compare them with our proposed vacua (see Sec. \ref{sec:introSLE}).

\bibliography{SLEbib}

\begin{thebibliography}{57}%
\makeatletter
\providecommand \@ifxundefined [1]{%
 \@ifx{#1\undefined}
}%
\providecommand \@ifnum [1]{%
 \ifnum #1\expandafter \@firstoftwo
 \else \expandafter \@secondoftwo
 \fi
}%
\providecommand \@ifx [1]{%
 \ifx #1\expandafter \@firstoftwo
 \else \expandafter \@secondoftwo
 \fi
}%
\providecommand \natexlab [1]{#1}%
\providecommand \enquote  [1]{``#1''}%
\providecommand \bibnamefont  [1]{#1}%
\providecommand \bibfnamefont [1]{#1}%
\providecommand \citenamefont [1]{#1}%
\providecommand \href@noop [0]{\@secondoftwo}%
\providecommand \href [0]{\begingroup \@sanitize@url \@href}%
\providecommand \@href[1]{\@@startlink{#1}\@@href}%
\providecommand \@@href[1]{\endgroup#1\@@endlink}%
\providecommand \@sanitize@url [0]{\catcode `\\12\catcode `\$12\catcode
  `\&12\catcode `\#12\catcode `\^12\catcode `\_12\catcode `\%12\relax}%
\providecommand \@@startlink[1]{}%
\providecommand \@@endlink[0]{}%
\providecommand \url  [0]{\begingroup\@sanitize@url \@url }%
\providecommand \@url [1]{\endgroup\@href {#1}{\urlprefix }}%
\providecommand \urlprefix  [0]{URL }%
\providecommand \Eprint [0]{\href }%
\providecommand \doibase [0]{http://dx.doi.org/}%
\providecommand \selectlanguage [0]{\@gobble}%
\providecommand \bibinfo  [0]{\@secondoftwo}%
\providecommand \bibfield  [0]{\@secondoftwo}%
\providecommand \translation [1]{[#1]}%
\providecommand \BibitemOpen [0]{}%
\providecommand \bibitemStop [0]{}%
\providecommand \bibitemNoStop [0]{.\EOS\space}%
\providecommand \EOS [0]{\spacefactor3000\relax}%
\providecommand \BibitemShut  [1]{\csname bibitem#1\endcsname}%
\let\auto@bib@innerbib\@empty
\bibitem [{\citenamefont {Aghanim}\ \emph {et~al.}(2020)\citenamefont {Aghanim}
  \emph {et~al.}}]{Planck2018_parameters}%
  \BibitemOpen
  \bibfield  {author} {\bibinfo {author} {\bibfnamefont {N.}~\bibnamefont
  {Aghanim}} \emph {et~al.} (\bibinfo {collaboration} {Planck}),\ }\href
  {\doibase 10.1051/0004-6361/201833910} {\bibfield  {journal} {\bibinfo
  {journal} {Astron. Astrophys.}\ }\textbf {\bibinfo {volume} {641}},\ \bibinfo
  {pages} {A6} (\bibinfo {year} {2020})},\ \Eprint
  {http://arxiv.org/abs/1807.06209} {arXiv:1807.06209 [astro-ph.CO]}
  \BibitemShut {NoStop}%
\bibitem [{\citenamefont {Akrami}\ \emph {et~al.}(2020)\citenamefont {Akrami}
  \emph {et~al.}}]{Planck2018_inflation}%
  \BibitemOpen
  \bibfield  {author} {\bibinfo {author} {\bibfnamefont {Y.}~\bibnamefont
  {Akrami}} \emph {et~al.} (\bibinfo {collaboration} {Planck}),\ }\href
  {\doibase 10.1051/0004-6361/201833887} {\bibfield  {journal} {\bibinfo
  {journal} {Astron. Astrophys.}\ }\textbf {\bibinfo {volume} {641}},\ \bibinfo
  {pages} {A10} (\bibinfo {year} {2020})},\ \Eprint
  {http://arxiv.org/abs/1807.06211} {arXiv:1807.06211 [astro-ph.CO]}
  \BibitemShut {NoStop}%
\bibitem [{\citenamefont {Parker}(1969)}]{parker1969}%
  \BibitemOpen
  \bibfield  {author} {\bibinfo {author} {\bibfnamefont {L.}~\bibnamefont
  {Parker}},\ }\href@noop {} {\bibfield  {journal} {\bibinfo  {journal}
  {Physical Review}\ }\textbf {\bibinfo {volume} {183}},\ \bibinfo {pages}
  {1057} (\bibinfo {year} {1969})}\BibitemShut {NoStop}%
\bibitem [{\citenamefont {Fulling}(1979)}]{fulling1979}%
  \BibitemOpen
  \bibfield  {author} {\bibinfo {author} {\bibfnamefont {S.}~\bibnamefont
  {Fulling}},\ }\href@noop {} {\bibfield  {journal} {\bibinfo  {journal}
  {General Relativity and Gravitation}\ }\textbf {\bibinfo {volume} {10}},\
  \bibinfo {pages} {807} (\bibinfo {year} {1979})}\BibitemShut {NoStop}%
\bibitem [{\citenamefont {Fahn}\ \emph {et~al.}(2019)\citenamefont {Fahn},
  \citenamefont {Giesel},\ and\ \citenamefont {Kobler}}]{Fahn:2018}%
  \BibitemOpen
  \bibfield  {author} {\bibinfo {author} {\bibfnamefont {M.~J.}\ \bibnamefont
  {Fahn}}, \bibinfo {author} {\bibfnamefont {K.}~\bibnamefont {Giesel}}, \ and\
  \bibinfo {author} {\bibfnamefont {M.}~\bibnamefont {Kobler}},\ }\href
  {\doibase 10.3390/universe5070170} {\bibfield  {journal} {\bibinfo  {journal}
  {Universe}\ }\textbf {\bibinfo {volume} {5}},\ \bibinfo {pages} {170}
  (\bibinfo {year} {2019})},\ \Eprint {http://arxiv.org/abs/1812.11122}
  {arXiv:1812.11122 [gr-qc]} \BibitemShut {NoStop}%
\bibitem [{\citenamefont {Elizaga~Navascu\'es}\ \emph
  {et~al.}(2019)\citenamefont {Elizaga~Navascu\'es}, \citenamefont
  {Mena~Marug\'an},\ and\ \citenamefont {Thiemann}}]{Elizaga2019}%
  \BibitemOpen
  \bibfield  {author} {\bibinfo {author} {\bibfnamefont {B.}~\bibnamefont
  {Elizaga~Navascu\'es}}, \bibinfo {author} {\bibfnamefont {G.~A.}\
  \bibnamefont {Mena~Marug\'an}}, \ and\ \bibinfo {author} {\bibfnamefont
  {T.}~\bibnamefont {Thiemann}},\ }\href {\doibase 10.1088/1361-6382/ab32af}
  {\bibfield  {journal} {\bibinfo  {journal} {Class. Quant. Grav.}\ }\textbf
  {\bibinfo {volume} {36}},\ \bibinfo {pages} {185010} (\bibinfo {year}
  {2019})},\ \Eprint {http://arxiv.org/abs/1903.05695} {arXiv:1903.05695
  [gr-qc]} \BibitemShut {NoStop}%
\bibitem [{\citenamefont {Agullo}\ \emph
  {et~al.}(2015{\natexlab{a}})\citenamefont {Agullo}, \citenamefont {Nelson},\
  and\ \citenamefont {Ashtekar}}]{agullo2015}%
  \BibitemOpen
  \bibfield  {author} {\bibinfo {author} {\bibfnamefont {I.}~\bibnamefont
  {Agullo}}, \bibinfo {author} {\bibfnamefont {W.}~\bibnamefont {Nelson}}, \
  and\ \bibinfo {author} {\bibfnamefont {A.}~\bibnamefont {Ashtekar}},\
  }\href@noop {} {\bibfield  {journal} {\bibinfo  {journal} {Physical Review
  D}\ }\textbf {\bibinfo {volume} {91}},\ \bibinfo {pages} {064051} (\bibinfo
  {year} {2015}{\natexlab{a}})},\ \Eprint {http://arxiv.org/abs/1412.3524}
  {arXiv:1412.3524 [gr-qc]} \BibitemShut {NoStop}%
\bibitem [{\citenamefont {Handley}\ \emph {et~al.}(2016)\citenamefont
  {Handley}, \citenamefont {Lasenby},\ and\ \citenamefont
  {Hobson}}]{handley2016}%
  \BibitemOpen
  \bibfield  {author} {\bibinfo {author} {\bibfnamefont {W.}~\bibnamefont
  {Handley}}, \bibinfo {author} {\bibfnamefont {A.}~\bibnamefont {Lasenby}}, \
  and\ \bibinfo {author} {\bibfnamefont {M.}~\bibnamefont {Hobson}},\
  }\href@noop {} {\bibfield  {journal} {\bibinfo  {journal} {Physical Review
  D}\ }\textbf {\bibinfo {volume} {94}},\ \bibinfo {pages} {024041} (\bibinfo
  {year} {2016})},\ \Eprint {http://arxiv.org/abs/1607.04148} {arXiv:1607.04148
  [gr-qc]} \BibitemShut {NoStop}%
\bibitem [{\citenamefont {Danielsson}(2002)}]{danielsson2002}%
  \BibitemOpen
  \bibfield  {author} {\bibinfo {author} {\bibfnamefont {U.~H.}\ \bibnamefont
  {Danielsson}},\ }\href@noop {} {\bibfield  {journal} {\bibinfo  {journal}
  {Physical Review D}\ }\textbf {\bibinfo {volume} {66}},\ \bibinfo {pages}
  {023511} (\bibinfo {year} {2002})},\ \Eprint
  {http://arxiv.org/abs/hep-th/0203198} {arXiv:hep-th/0203198} \BibitemShut
  {NoStop}%
\bibitem [{\citenamefont {Ashtekar}\ and\ \citenamefont
  {Gupt}(2017)}]{Ashtekar:2016}%
  \BibitemOpen
  \bibfield  {author} {\bibinfo {author} {\bibfnamefont {A.}~\bibnamefont
  {Ashtekar}}\ and\ \bibinfo {author} {\bibfnamefont {B.}~\bibnamefont
  {Gupt}},\ }\href {\doibase 10.1088/1361-6382/aa52d4} {\bibfield  {journal}
  {\bibinfo  {journal} {Class. Quant. Grav.}\ }\textbf {\bibinfo {volume}
  {34}},\ \bibinfo {pages} {035004} (\bibinfo {year} {2017})},\ \Eprint
  {http://arxiv.org/abs/1610.09424} {arXiv:1610.09424 [gr-qc]} \BibitemShut
  {NoStop}%
\bibitem [{\citenamefont {de~Blas}\ and\ \citenamefont
  {Olmedo}(2016)}]{nonosc}%
  \BibitemOpen
  \bibfield  {author} {\bibinfo {author} {\bibfnamefont {D.~M.}\ \bibnamefont
  {de~Blas}}\ and\ \bibinfo {author} {\bibfnamefont {J.}~\bibnamefont
  {Olmedo}},\ }\href {\doibase 10.1088/1475-7516/2016/06/029} {\bibfield
  {journal} {\bibinfo  {journal} {JCAP}\ }\textbf {\bibinfo {volume} {06}},\
  \bibinfo {pages} {029} (\bibinfo {year} {2016})},\ \Eprint
  {http://arxiv.org/abs/1601.01716} {arXiv:1601.01716 [gr-qc]} \BibitemShut
  {NoStop}%
\bibitem [{\citenamefont {Elizaga~Navascu\'es}\ \emph
  {et~al.}(2020)\citenamefont {Elizaga~Navascu\'es}, \citenamefont
  {Marug\'an},\ and\ \citenamefont {Prado}}]{menava}%
  \BibitemOpen
  \bibfield  {author} {\bibinfo {author} {\bibfnamefont {B.}~\bibnamefont
  {Elizaga~Navascu\'es}}, \bibinfo {author} {\bibfnamefont {G.~A.~M.}\
  \bibnamefont {Marug\'an}}, \ and\ \bibinfo {author} {\bibfnamefont
  {S.}~\bibnamefont {Prado}},\ }\href {\doibase 10.1088/1361-6382/abc6bb}
  {\bibfield  {journal} {\bibinfo  {journal} {Class. Quant. Grav.}\ }\textbf
  {\bibinfo {volume} {38}},\ \bibinfo {pages} {035001} (\bibinfo {year}
  {2020})},\ \Eprint {http://arxiv.org/abs/2005.10194} {arXiv:2005.10194
  [gr-qc]} \BibitemShut {NoStop}%
\bibitem [{\citenamefont {Navascu\'es}\ and\ \citenamefont
  {Mena~Marug\'an}(2021)}]{Navascues:2021}%
  \BibitemOpen
  \bibfield  {author} {\bibinfo {author} {\bibfnamefont {B.~E.}\ \bibnamefont
  {Navascu\'es}}\ and\ \bibinfo {author} {\bibfnamefont {G.~A.}\ \bibnamefont
  {Mena~Marug\'an}},\ }\href@noop {} {\  (\bibinfo {year} {2021})},\ \Eprint
  {http://arxiv.org/abs/2104.15002} {arXiv:2104.15002 [gr-qc]} \BibitemShut
  {NoStop}%
\bibitem [{\citenamefont {Olbermann}(2007)}]{Olbermann2007}%
  \BibitemOpen
  \bibfield  {author} {\bibinfo {author} {\bibfnamefont {H.}~\bibnamefont
  {Olbermann}},\ }\href {\doibase 10.1088/0264-9381/24/20/007} {\bibfield
  {journal} {\bibinfo  {journal} {Class. Quant. Grav.}\ }\textbf {\bibinfo
  {volume} {24}},\ \bibinfo {pages} {5011} (\bibinfo {year} {2007})},\ \Eprint
  {http://arxiv.org/abs/0704.2986} {arXiv:0704.2986 [gr-qc]} \BibitemShut
  {NoStop}%
\bibitem [{\citenamefont {Fewster}(2000)}]{Fewster2000}%
  \BibitemOpen
  \bibfield  {author} {\bibinfo {author} {\bibfnamefont {C.~J.}\ \bibnamefont
  {Fewster}},\ }\href {\doibase 10.1088/0264-9381/17/9/302} {\bibfield
  {journal} {\bibinfo  {journal} {Classical and Quantum Gravity}\ }\textbf
  {\bibinfo {volume} {17}},\ \bibinfo {pages} {1897–1911} (\bibinfo {year}
  {2000})},\ \Eprint {http://arxiv.org/abs/gr-qc/9910060} {arXiv:gr-qc/9910060}
  \BibitemShut {NoStop}%
\bibitem [{\citenamefont {Afshordi}\ \emph {et~al.}(2012)\citenamefont
  {Afshordi}, \citenamefont {Aslanbeigi},\ and\ \citenamefont
  {Sorkin}}]{afshordi2012}%
  \BibitemOpen
  \bibfield  {author} {\bibinfo {author} {\bibfnamefont {N.}~\bibnamefont
  {Afshordi}}, \bibinfo {author} {\bibfnamefont {S.}~\bibnamefont
  {Aslanbeigi}}, \ and\ \bibinfo {author} {\bibfnamefont {R.~D.}\ \bibnamefont
  {Sorkin}},\ }\href@noop {} {\bibfield  {journal} {\bibinfo  {journal}
  {Journal of High Energy Physics}\ }\textbf {\bibinfo {volume} {2012}},\
  \bibinfo {pages} {137} (\bibinfo {year} {2012})},\ \Eprint
  {http://arxiv.org/abs/1205.1296} {arXiv:1205.1296 [hep-th]} \BibitemShut
  {NoStop}%
\bibitem [{\citenamefont {Fewster}\ and\ \citenamefont
  {Verch}(2012)}]{Fewster_2012}%
  \BibitemOpen
  \bibfield  {author} {\bibinfo {author} {\bibfnamefont {C.~J.}\ \bibnamefont
  {Fewster}}\ and\ \bibinfo {author} {\bibfnamefont {R.}~\bibnamefont
  {Verch}},\ }\href {\doibase 10.1088/0264-9381/29/20/205017} {\bibfield
  {journal} {\bibinfo  {journal} {Classical and Quantum Gravity}\ }\textbf
  {\bibinfo {volume} {29}},\ \bibinfo {pages} {205017} (\bibinfo {year}
  {2012})},\ \Eprint {http://arxiv.org/abs/1206.1562} {arXiv:1206.1562
  [math-ph]} \BibitemShut {NoStop}%
\bibitem [{\citenamefont {Degner}\ and\ \citenamefont
  {DESY}(2013)}]{DegnerPhD2013}%
  \BibitemOpen
  \bibfield  {author} {\bibinfo {author} {\bibfnamefont {A.}~\bibnamefont
  {Degner}}\ and\ \bibinfo {author} {\bibnamefont {DESY}},\ }\emph {\bibinfo
  {title} {{P}roperties of {S}tates of {L}ow {E}nergy on {C}osmological
  {S}pacetimes}},\ \href {\doibase 10.3204/DESY-THESIS-2013-002} {\bibinfo
  {type} {Dr.}},\ \bibinfo  {school} {Universit\"at Hamburg} (\bibinfo {year}
  {2013}),\ \bibinfo {note} {universit\"at Hamburg, Diss., 2013}\BibitemShut
  {NoStop}%
\bibitem [{\citenamefont {Banerjee}\ and\ \citenamefont
  {Niedermaier}(2020)}]{Niedermaier2020}%
  \BibitemOpen
  \bibfield  {author} {\bibinfo {author} {\bibfnamefont {R.}~\bibnamefont
  {Banerjee}}\ and\ \bibinfo {author} {\bibfnamefont {M.}~\bibnamefont
  {Niedermaier}},\ }\href {\doibase 10.1063/5.0019311} {\bibfield  {journal}
  {\bibinfo  {journal} {J. Math. Phys.}\ }\textbf {\bibinfo {volume} {61}},\
  \bibinfo {pages} {103511} (\bibinfo {year} {2020})},\ \Eprint
  {http://arxiv.org/abs/2006.08685} {arXiv:2006.08685 [math-ph]} \BibitemShut
  {NoStop}%
\bibitem [{\citenamefont {Bojowald}(2005)}]{LQCreview_Bojowald2005}%
  \BibitemOpen
  \bibfield  {author} {\bibinfo {author} {\bibfnamefont {M.}~\bibnamefont
  {Bojowald}},\ }\href@noop {} {\bibfield  {journal} {\bibinfo  {journal}
  {Living Rev. Rel.}\ }\textbf {\bibinfo {volume} {8}},\ \bibinfo {pages} {11}
  (\bibinfo {year} {2005})},\ \Eprint {http://arxiv.org/abs/gr-qc/0601085}
  {arXiv:gr-qc/0601085 [gr-qc]} \BibitemShut {NoStop}%
\bibitem [{\citenamefont {Ashtekar}\ and\ \citenamefont
  {Singh}(2011)}]{LQCreview_Ashtekar2011}%
  \BibitemOpen
  \bibfield  {author} {\bibinfo {author} {\bibfnamefont {A.}~\bibnamefont
  {Ashtekar}}\ and\ \bibinfo {author} {\bibfnamefont {P.}~\bibnamefont
  {Singh}},\ }\href@noop {} {\bibfield  {journal} {\bibinfo  {journal} {Class.
  Quant. Grav.}\ }\textbf {\bibinfo {volume} {28}},\ \bibinfo {pages} {213001}
  (\bibinfo {year} {2011})},\ \Eprint {http://arxiv.org/abs/1108.0893}
  {arXiv:1108.0893 [gr-qc]} \BibitemShut {NoStop}%
\bibitem [{\citenamefont {Banerjee}\ \emph {et~al.}(2012)\citenamefont
  {Banerjee}, \citenamefont {Calcagni},\ and\ \citenamefont
  {Martin-Benito}}]{LQCreview_Banerjee2012}%
  \BibitemOpen
  \bibfield  {author} {\bibinfo {author} {\bibfnamefont {K.}~\bibnamefont
  {Banerjee}}, \bibinfo {author} {\bibfnamefont {G.}~\bibnamefont {Calcagni}},
  \ and\ \bibinfo {author} {\bibfnamefont {M.}~\bibnamefont {Martin-Benito}},\
  }\href@noop {} {\bibfield  {journal} {\bibinfo  {journal} {SIGMA}\ }\textbf
  {\bibinfo {volume} {8}},\ \bibinfo {pages} {016} (\bibinfo {year} {2012})},\
  \Eprint {http://arxiv.org/abs/1109.6801} {arXiv:1109.6801 [gr-qc]}
  \BibitemShut {NoStop}%
\bibitem [{\citenamefont {Agullo}\ and\ \citenamefont
  {Singh}(2017)}]{LQCreview_Agullo2016}%
  \BibitemOpen
  \bibfield  {author} {\bibinfo {author} {\bibfnamefont {I.}~\bibnamefont
  {Agullo}}\ and\ \bibinfo {author} {\bibfnamefont {P.}~\bibnamefont {Singh}},\
  }in\ \href@noop {} {\emph {\bibinfo {booktitle} {Loop Quantum Gravity: The
  First 30 Years}}},\ \bibinfo {editor} {edited by\ \bibinfo {editor}
  {\bibfnamefont {A.}~\bibnamefont {Ashtekar}}\ and\ \bibinfo {editor}
  {\bibfnamefont {J.}~\bibnamefont {Pullin}}}\ (\bibinfo  {publisher} {WSP},\
  \bibinfo {year} {2017})\ pp.\ \bibinfo {pages} {183--240},\ \Eprint
  {http://arxiv.org/abs/1612.01236} {arXiv:1612.01236 [gr-qc]} \BibitemShut
  {NoStop}%
\bibitem [{\citenamefont {Ashtekar}\ \emph
  {et~al.}(2006{\natexlab{a}})\citenamefont {Ashtekar}, \citenamefont
  {Pawlowski},\ and\ \citenamefont {Singh}}]{APS_PRL}%
  \BibitemOpen
  \bibfield  {author} {\bibinfo {author} {\bibfnamefont {A.}~\bibnamefont
  {Ashtekar}}, \bibinfo {author} {\bibfnamefont {T.}~\bibnamefont {Pawlowski}},
  \ and\ \bibinfo {author} {\bibfnamefont {P.}~\bibnamefont {Singh}},\
  }\href@noop {} {\bibfield  {journal} {\bibinfo  {journal} {Phys. Rev. Lett.}\
  }\textbf {\bibinfo {volume} {96}},\ \bibinfo {pages} {141301} (\bibinfo
  {year} {2006}{\natexlab{a}})},\ \Eprint {http://arxiv.org/abs/gr-qc/0602086}
  {arXiv:gr-qc/0602086 [gr-qc]} \BibitemShut {NoStop}%
\bibitem [{\citenamefont {Ashtekar}\ \emph
  {et~al.}(2006{\natexlab{b}})\citenamefont {Ashtekar}, \citenamefont
  {Pawlowski},\ and\ \citenamefont {Singh}}]{APS_extended}%
  \BibitemOpen
  \bibfield  {author} {\bibinfo {author} {\bibfnamefont {A.}~\bibnamefont
  {Ashtekar}}, \bibinfo {author} {\bibfnamefont {T.}~\bibnamefont {Pawlowski}},
  \ and\ \bibinfo {author} {\bibfnamefont {P.}~\bibnamefont {Singh}},\
  }\href@noop {} {\bibfield  {journal} {\bibinfo  {journal} {Phys. Rev.}\
  }\textbf {\bibinfo {volume} {D74}},\ \bibinfo {pages} {084003} (\bibinfo
  {year} {2006}{\natexlab{b}})},\ \Eprint {http://arxiv.org/abs/gr-qc/0607039}
  {arXiv:gr-qc/0607039 [gr-qc]} \BibitemShut {NoStop}%
\bibitem [{\citenamefont {Bentivegna}\ and\ \citenamefont
  {Pawlowski}(2008)}]{Bentivegna2008}%
  \BibitemOpen
  \bibfield  {author} {\bibinfo {author} {\bibfnamefont {E.}~\bibnamefont
  {Bentivegna}}\ and\ \bibinfo {author} {\bibfnamefont {T.}~\bibnamefont
  {Pawlowski}},\ }\href@noop {} {\bibfield  {journal} {\bibinfo  {journal}
  {Phys. Rev.}\ }\textbf {\bibinfo {volume} {D77}},\ \bibinfo {pages} {124025}
  (\bibinfo {year} {2008})},\ \Eprint {http://arxiv.org/abs/0803.4446}
  {arXiv:0803.4446 [gr-qc]} \BibitemShut {NoStop}%
\bibitem [{\citenamefont {Kaminski}\ and\ \citenamefont
  {Pawlowski}(2010)}]{Kaminski2009}%
  \BibitemOpen
  \bibfield  {author} {\bibinfo {author} {\bibfnamefont {W.}~\bibnamefont
  {Kaminski}}\ and\ \bibinfo {author} {\bibfnamefont {T.}~\bibnamefont
  {Pawlowski}},\ }\href@noop {} {\bibfield  {journal} {\bibinfo  {journal}
  {Phys. Rev.}\ }\textbf {\bibinfo {volume} {D81}},\ \bibinfo {pages} {024014}
  (\bibinfo {year} {2010})},\ \Eprint {http://arxiv.org/abs/0912.0162}
  {arXiv:0912.0162 [gr-qc]} \BibitemShut {NoStop}%
\bibitem [{\citenamefont {Pawlowski}\ and\ \citenamefont
  {Ashtekar}(2012)}]{Pawlowski2012}%
  \BibitemOpen
  \bibfield  {author} {\bibinfo {author} {\bibfnamefont {T.}~\bibnamefont
  {Pawlowski}}\ and\ \bibinfo {author} {\bibfnamefont {A.}~\bibnamefont
  {Ashtekar}},\ }\href@noop {} {\bibfield  {journal} {\bibinfo  {journal}
  {Phys. Rev.}\ }\textbf {\bibinfo {volume} {D85}},\ \bibinfo {pages} {064001}
  (\bibinfo {year} {2012})},\ \Eprint {http://arxiv.org/abs/1112.0360}
  {arXiv:1112.0360 [gr-qc]} \BibitemShut {NoStop}%
\bibitem [{\citenamefont {Fernandez-Mendez}\ \emph {et~al.}(2012)\citenamefont
  {Fernandez-Mendez}, \citenamefont {Mena~Marugan},\ and\ \citenamefont
  {Olmedo}}]{hybrid_FernandezMendez2012}%
  \BibitemOpen
  \bibfield  {author} {\bibinfo {author} {\bibfnamefont {M.}~\bibnamefont
  {Fernandez-Mendez}}, \bibinfo {author} {\bibfnamefont {G.~A.}\ \bibnamefont
  {Mena~Marugan}}, \ and\ \bibinfo {author} {\bibfnamefont {J.}~\bibnamefont
  {Olmedo}},\ }\href {\doibase 10.1103/PhysRevD.86.024003} {\bibfield
  {journal} {\bibinfo  {journal} {Phys. Rev. D}\ }\textbf {\bibinfo {volume}
  {86}},\ \bibinfo {pages} {024003} (\bibinfo {year} {2012})},\ \Eprint
  {http://arxiv.org/abs/1205.1917} {arXiv:1205.1917 [gr-qc]} \BibitemShut
  {NoStop}%
\bibitem [{\citenamefont {Fern\'andez-M\'endez}\ \emph
  {et~al.}(2013)\citenamefont {Fern\'andez-M\'endez}, \citenamefont
  {Mena~Marug\'an},\ and\ \citenamefont {Olmedo}}]{hybrid_FernandezMendez2013}%
  \BibitemOpen
  \bibfield  {author} {\bibinfo {author} {\bibfnamefont {M.}~\bibnamefont
  {Fern\'andez-M\'endez}}, \bibinfo {author} {\bibfnamefont {G.~A.}\
  \bibnamefont {Mena~Marug\'an}}, \ and\ \bibinfo {author} {\bibfnamefont
  {J.}~\bibnamefont {Olmedo}},\ }\href {\doibase 10.1103/PhysRevD.88.044013}
  {\bibfield  {journal} {\bibinfo  {journal} {Phys. Rev. D}\ }\textbf {\bibinfo
  {volume} {88}},\ \bibinfo {pages} {044013} (\bibinfo {year} {2013})},\
  \Eprint {http://arxiv.org/abs/1307.5222} {arXiv:1307.5222 [gr-qc]}
  \BibitemShut {NoStop}%
\bibitem [{\citenamefont {Fern\'andez-M\'endez}\ \emph
  {et~al.}(2014)\citenamefont {Fern\'andez-M\'endez}, \citenamefont
  {Mena~Marug\'an},\ and\ \citenamefont {Olmedo}}]{hybrid_FernandezMendez2014}%
  \BibitemOpen
  \bibfield  {author} {\bibinfo {author} {\bibfnamefont {M.}~\bibnamefont
  {Fern\'andez-M\'endez}}, \bibinfo {author} {\bibfnamefont {G.~A.}\
  \bibnamefont {Mena~Marug\'an}}, \ and\ \bibinfo {author} {\bibfnamefont
  {J.}~\bibnamefont {Olmedo}},\ }\href {\doibase 10.1103/PhysRevD.89.044041}
  {\bibfield  {journal} {\bibinfo  {journal} {Phys. Rev. D}\ }\textbf {\bibinfo
  {volume} {89}},\ \bibinfo {pages} {044041} (\bibinfo {year} {2014})},\
  \Eprint {http://arxiv.org/abs/1401.5256} {arXiv:1401.5256 [gr-qc]}
  \BibitemShut {NoStop}%
\bibitem [{\citenamefont {Gomar}\ \emph {et~al.}(2014)\citenamefont {Gomar},
  \citenamefont {Fern\'andez-M\'endez}, \citenamefont {Marug\'an},\ and\
  \citenamefont {Olmedo}}]{hybrid_Gomar2014}%
  \BibitemOpen
  \bibfield  {author} {\bibinfo {author} {\bibfnamefont {L.~C.}\ \bibnamefont
  {Gomar}}, \bibinfo {author} {\bibfnamefont {M.}~\bibnamefont
  {Fern\'andez-M\'endez}}, \bibinfo {author} {\bibfnamefont {G.~A.~M.}\
  \bibnamefont {Marug\'an}}, \ and\ \bibinfo {author} {\bibfnamefont
  {J.}~\bibnamefont {Olmedo}},\ }\href {\doibase 10.1103/PhysRevD.90.064015}
  {\bibfield  {journal} {\bibinfo  {journal} {Phys. Rev. D}\ }\textbf {\bibinfo
  {volume} {90}},\ \bibinfo {pages} {064015} (\bibinfo {year} {2014})},\
  \Eprint {http://arxiv.org/abs/1407.0998} {arXiv:1407.0998 [gr-qc]}
  \BibitemShut {NoStop}%
\bibitem [{\citenamefont {Gomar}\ \emph {et~al.}(2015)\citenamefont {Gomar},
  \citenamefont {Mart\'\i{}n-Benito},\ and\ \citenamefont
  {Marug\'an}}]{hybrid_Gomar2015}%
  \BibitemOpen
  \bibfield  {author} {\bibinfo {author} {\bibfnamefont {L.~C.}\ \bibnamefont
  {Gomar}}, \bibinfo {author} {\bibfnamefont {M.}~\bibnamefont
  {Mart\'\i{}n-Benito}}, \ and\ \bibinfo {author} {\bibfnamefont {G.~A.~M.}\
  \bibnamefont {Marug\'an}},\ }\href {\doibase 10.1088/1475-7516/2015/06/045}
  {\bibfield  {journal} {\bibinfo  {journal} {JCAP}\ }\textbf {\bibinfo
  {volume} {06}},\ \bibinfo {pages} {045} (\bibinfo {year} {2015})},\ \Eprint
  {http://arxiv.org/abs/1503.03907} {arXiv:1503.03907 [gr-qc]} \BibitemShut
  {NoStop}%
\bibitem [{\citenamefont {Castell\'o~Gomar}\ \emph {et~al.}(2016)\citenamefont
  {Castell\'o~Gomar}, \citenamefont {Mart\'\i{}n-Benito},\ and\ \citenamefont
  {Mena~Marug\'an}}]{hyb-ms}%
  \BibitemOpen
  \bibfield  {author} {\bibinfo {author} {\bibfnamefont {L.}~\bibnamefont
  {Castell\'o~Gomar}}, \bibinfo {author} {\bibfnamefont {M.}~\bibnamefont
  {Mart\'\i{}n-Benito}}, \ and\ \bibinfo {author} {\bibfnamefont {G.~A.}\
  \bibnamefont {Mena~Marug\'an}},\ }\href {\doibase 10.1103/PhysRevD.93.104025}
  {\bibfield  {journal} {\bibinfo  {journal} {Phys. Rev. D}\ }\textbf {\bibinfo
  {volume} {93}},\ \bibinfo {pages} {104025} (\bibinfo {year} {2016})},\
  \Eprint {http://arxiv.org/abs/1603.08448} {arXiv:1603.08448 [gr-qc]}
  \BibitemShut {NoStop}%
\bibitem [{\citenamefont {Mart\'\i{}nez}\ and\ \citenamefont
  {Olmedo}(2016)}]{hybrid_Martinez2016}%
  \BibitemOpen
  \bibfield  {author} {\bibinfo {author} {\bibfnamefont {F.~B.}\ \bibnamefont
  {Mart\'\i{}nez}}\ and\ \bibinfo {author} {\bibfnamefont {J.}~\bibnamefont
  {Olmedo}},\ }\href {\doibase 10.1103/PhysRevD.93.124008} {\bibfield
  {journal} {\bibinfo  {journal} {Phys. Rev. D}\ }\textbf {\bibinfo {volume}
  {93}},\ \bibinfo {pages} {124008} (\bibinfo {year} {2016})},\ \Eprint
  {http://arxiv.org/abs/1605.04293} {arXiv:1605.04293 [gr-qc]} \BibitemShut
  {NoStop}%
\bibitem [{\citenamefont {Elizaga~Navascu\'es}\ \emph
  {et~al.}(2018)\citenamefont {Elizaga~Navascu\'es}, \citenamefont {Martin~de
  Blas},\ and\ \citenamefont {Mena~Marug\'an}}]{hyb-vs-dress}%
  \BibitemOpen
  \bibfield  {author} {\bibinfo {author} {\bibfnamefont {B.}~\bibnamefont
  {Elizaga~Navascu\'es}}, \bibinfo {author} {\bibfnamefont {D.}~\bibnamefont
  {Martin~de Blas}}, \ and\ \bibinfo {author} {\bibfnamefont {G.~A.}\
  \bibnamefont {Mena~Marug\'an}},\ }\href {\doibase 10.1103/PhysRevD.97.043523}
  {\bibfield  {journal} {\bibinfo  {journal} {Phys. Rev. D}\ }\textbf {\bibinfo
  {volume} {97}},\ \bibinfo {pages} {043523} (\bibinfo {year} {2018})},\
  \Eprint {http://arxiv.org/abs/1711.10861} {arXiv:1711.10861 [gr-qc]}
  \BibitemShut {NoStop}%
\bibitem [{\citenamefont {Castell\'o~Gomar}\ \emph {et~al.}(2017)\citenamefont
  {Castell\'o~Gomar}, \citenamefont {Mena~Marug\'an}, \citenamefont
  {Mart\'\i{}n De~Blas},\ and\ \citenamefont {Olmedo}}]{hyb-obs}%
  \BibitemOpen
  \bibfield  {author} {\bibinfo {author} {\bibfnamefont {L.}~\bibnamefont
  {Castell\'o~Gomar}}, \bibinfo {author} {\bibfnamefont {G.~A.}\ \bibnamefont
  {Mena~Marug\'an}}, \bibinfo {author} {\bibfnamefont {D.}~\bibnamefont
  {Mart\'\i{}n De~Blas}}, \ and\ \bibinfo {author} {\bibfnamefont
  {J.}~\bibnamefont {Olmedo}},\ }\href {\doibase 10.1103/PhysRevD.96.103528}
  {\bibfield  {journal} {\bibinfo  {journal} {Phys. Rev. D}\ }\textbf {\bibinfo
  {volume} {96}},\ \bibinfo {pages} {103528} (\bibinfo {year} {2017})},\
  \Eprint {http://arxiv.org/abs/1702.06036} {arXiv:1702.06036 [gr-qc]}
  \BibitemShut {NoStop}%
\bibitem [{\citenamefont {Agullo}\ \emph {et~al.}(2012)\citenamefont {Agullo},
  \citenamefont {Ashtekar},\ and\ \citenamefont
  {Nelson}}]{dressedmetric_Agullo2012}%
  \BibitemOpen
  \bibfield  {author} {\bibinfo {author} {\bibfnamefont {I.}~\bibnamefont
  {Agullo}}, \bibinfo {author} {\bibfnamefont {A.}~\bibnamefont {Ashtekar}}, \
  and\ \bibinfo {author} {\bibfnamefont {W.}~\bibnamefont {Nelson}},\ }\href
  {\doibase 10.1103/PhysRevLett.109.251301} {\bibfield  {journal} {\bibinfo
  {journal} {Phys. Rev. Lett.}\ }\textbf {\bibinfo {volume} {109}},\ \bibinfo
  {pages} {251301} (\bibinfo {year} {2012})},\ \Eprint
  {http://arxiv.org/abs/1209.1609} {arXiv:1209.1609 [gr-qc]} \BibitemShut
  {NoStop}%
\bibitem [{\citenamefont {Agullo}\ \emph
  {et~al.}(2013{\natexlab{a}})\citenamefont {Agullo}, \citenamefont
  {Ashtekar},\ and\ \citenamefont {Nelson}}]{dressedmetric_Agullo_PRD_2013}%
  \BibitemOpen
  \bibfield  {author} {\bibinfo {author} {\bibfnamefont {I.}~\bibnamefont
  {Agullo}}, \bibinfo {author} {\bibfnamefont {A.}~\bibnamefont {Ashtekar}}, \
  and\ \bibinfo {author} {\bibfnamefont {W.}~\bibnamefont {Nelson}},\ }\href
  {\doibase 10.1103/PhysRevD.87.043507} {\bibfield  {journal} {\bibinfo
  {journal} {Phys. Rev. D}\ }\textbf {\bibinfo {volume} {87}},\ \bibinfo
  {pages} {043507} (\bibinfo {year} {2013}{\natexlab{a}})},\ \Eprint
  {http://arxiv.org/abs/1211.1354} {arXiv:1211.1354 [gr-qc]} \BibitemShut
  {NoStop}%
\bibitem [{\citenamefont {Agullo}\ \emph
  {et~al.}(2013{\natexlab{b}})\citenamefont {Agullo}, \citenamefont
  {Ashtekar},\ and\ \citenamefont {Nelson}}]{dressedmetric_Agullo_CQG_2013}%
  \BibitemOpen
  \bibfield  {author} {\bibinfo {author} {\bibfnamefont {I.}~\bibnamefont
  {Agullo}}, \bibinfo {author} {\bibfnamefont {A.}~\bibnamefont {Ashtekar}}, \
  and\ \bibinfo {author} {\bibfnamefont {W.}~\bibnamefont {Nelson}},\ }\href
  {\doibase 10.1088/0264-9381/30/8/085014} {\bibfield  {journal} {\bibinfo
  {journal} {Classical and Quantum Gravity}\ }\textbf {\bibinfo {volume}
  {30}},\ \bibinfo {pages} {085014} (\bibinfo {year} {2013}{\natexlab{b}})},\
  \Eprint {http://arxiv.org/abs/1302.0254} {arXiv:1302.0254 [gr-qc]}
  \BibitemShut {NoStop}%
\bibitem [{\citenamefont {Agullo}\ \emph
  {et~al.}(2015{\natexlab{b}})\citenamefont {Agullo}, \citenamefont {Nelson},\
  and\ \citenamefont {Ashtekar}}]{instvac}%
  \BibitemOpen
  \bibfield  {author} {\bibinfo {author} {\bibfnamefont {I.}~\bibnamefont
  {Agullo}}, \bibinfo {author} {\bibfnamefont {W.}~\bibnamefont {Nelson}}, \
  and\ \bibinfo {author} {\bibfnamefont {A.}~\bibnamefont {Ashtekar}},\ }\href
  {\doibase 10.1103/PhysRevD.91.064051} {\bibfield  {journal} {\bibinfo
  {journal} {Phys. Rev. D}\ }\textbf {\bibinfo {volume} {91}},\ \bibinfo
  {pages} {064051} (\bibinfo {year} {2015}{\natexlab{b}})},\ \Eprint
  {http://arxiv.org/abs/1412.3524} {arXiv:1412.3524 [gr-qc]} \BibitemShut
  {NoStop}%
\bibitem [{\citenamefont {Mart\'\i{}n-Benito}\ \emph
  {et~al.}(2021)\citenamefont {Mart\'\i{}n-Benito}, \citenamefont {Neves},\
  and\ \citenamefont {Olmedo}}]{mno-sles}%
  \BibitemOpen
  \bibfield  {author} {\bibinfo {author} {\bibfnamefont {M.}~\bibnamefont
  {Mart\'\i{}n-Benito}}, \bibinfo {author} {\bibfnamefont {R.~B.}\ \bibnamefont
  {Neves}}, \ and\ \bibinfo {author} {\bibfnamefont {J.}~\bibnamefont
  {Olmedo}},\ }\href@noop {} {\  (\bibinfo {year} {2021})},\ \Eprint
  {http://arxiv.org/abs/2104.14850} {arXiv:2104.14850 [gr-qc]} \BibitemShut
  {NoStop}%
\bibitem [{\citenamefont {Ashtekar}\ and\ \citenamefont
  {Sloan}(2011)}]{AshtekarSloan_probinflation}%
  \BibitemOpen
  \bibfield  {author} {\bibinfo {author} {\bibfnamefont {A.}~\bibnamefont
  {Ashtekar}}\ and\ \bibinfo {author} {\bibfnamefont {D.}~\bibnamefont
  {Sloan}},\ }\href {\doibase 10.1007/s10714-011-1246-y} {\bibfield  {journal}
  {\bibinfo  {journal} {Gen. Rel. Grav.}\ }\textbf {\bibinfo {volume} {43}},\
  \bibinfo {pages} {3619} (\bibinfo {year} {2011})},\ \Eprint
  {http://arxiv.org/abs/1103.2475} {arXiv:1103.2475 [gr-qc]} \BibitemShut
  {NoStop}%
\bibitem [{\citenamefont {Dapor}\ and\ \citenamefont
  {Liegener}(2018)}]{Dapor:2017}%
  \BibitemOpen
  \bibfield  {author} {\bibinfo {author} {\bibfnamefont {A.}~\bibnamefont
  {Dapor}}\ and\ \bibinfo {author} {\bibfnamefont {K.}~\bibnamefont
  {Liegener}},\ }\href {\doibase 10.1016/j.physletb.2018.09.005} {\bibfield
  {journal} {\bibinfo  {journal} {Phys. Lett. B}\ }\textbf {\bibinfo {volume}
  {785}},\ \bibinfo {pages} {506} (\bibinfo {year} {2018})},\ \Eprint
  {http://arxiv.org/abs/1706.09833} {arXiv:1706.09833 [gr-qc]} \BibitemShut
  {NoStop}%
\bibitem [{\citenamefont {Assanioussi}\ \emph {et~al.}(2018)\citenamefont
  {Assanioussi}, \citenamefont {Dapor}, \citenamefont {Liegener},\ and\
  \citenamefont {Paw\l{}owski}}]{Assanioussi:2018}%
  \BibitemOpen
  \bibfield  {author} {\bibinfo {author} {\bibfnamefont {M.}~\bibnamefont
  {Assanioussi}}, \bibinfo {author} {\bibfnamefont {A.}~\bibnamefont {Dapor}},
  \bibinfo {author} {\bibfnamefont {K.}~\bibnamefont {Liegener}}, \ and\
  \bibinfo {author} {\bibfnamefont {T.}~\bibnamefont {Paw\l{}owski}},\ }\href
  {\doibase 10.1103/PhysRevLett.121.081303} {\bibfield  {journal} {\bibinfo
  {journal} {Phys. Rev. Lett.}\ }\textbf {\bibinfo {volume} {121}},\ \bibinfo
  {pages} {081303} (\bibinfo {year} {2018})},\ \Eprint
  {http://arxiv.org/abs/1801.00768} {arXiv:1801.00768 [gr-qc]} \BibitemShut
  {NoStop}%
\bibitem [{\citenamefont {Assanioussi}\ \emph {et~al.}(2019)\citenamefont
  {Assanioussi}, \citenamefont {Dapor}, \citenamefont {Liegener},\ and\
  \citenamefont {Paw\l{}owski}}]{Assanioussi:2019}%
  \BibitemOpen
  \bibfield  {author} {\bibinfo {author} {\bibfnamefont {M.}~\bibnamefont
  {Assanioussi}}, \bibinfo {author} {\bibfnamefont {A.}~\bibnamefont {Dapor}},
  \bibinfo {author} {\bibfnamefont {K.}~\bibnamefont {Liegener}}, \ and\
  \bibinfo {author} {\bibfnamefont {T.}~\bibnamefont {Paw\l{}owski}},\ }\href
  {\doibase 10.1103/PhysRevD.100.084003} {\bibfield  {journal} {\bibinfo
  {journal} {Phys. Rev. D}\ }\textbf {\bibinfo {volume} {100}},\ \bibinfo
  {pages} {084003} (\bibinfo {year} {2019})},\ \Eprint
  {http://arxiv.org/abs/1906.05315} {arXiv:1906.05315 [gr-qc]} \BibitemShut
  {NoStop}%
\bibitem [{\citenamefont {Garc\'\i{}a-Quismondo}\ and\ \citenamefont
  {Mena~Marug\'an}(2019)}]{quismondo:2019}%
  \BibitemOpen
  \bibfield  {author} {\bibinfo {author} {\bibfnamefont {A.}~\bibnamefont
  {Garc\'\i{}a-Quismondo}}\ and\ \bibinfo {author} {\bibfnamefont {G.~A.}\
  \bibnamefont {Mena~Marug\'an}},\ }\href {\doibase 10.1103/PhysRevD.99.083505}
  {\bibfield  {journal} {\bibinfo  {journal} {Phys. Rev. D}\ }\textbf {\bibinfo
  {volume} {99}},\ \bibinfo {pages} {083505} (\bibinfo {year} {2019})},\
  \Eprint {http://arxiv.org/abs/1903.00265} {arXiv:1903.00265 [gr-qc]}
  \BibitemShut {NoStop}%
\bibitem [{\citenamefont {Agullo}(2018)}]{Agullo:2018}%
  \BibitemOpen
  \bibfield  {author} {\bibinfo {author} {\bibfnamefont {I.}~\bibnamefont
  {Agullo}},\ }\href {\doibase 10.1007/s10714-018-2413-1} {\bibfield  {journal}
  {\bibinfo  {journal} {Gen. Rel. Grav.}\ }\textbf {\bibinfo {volume} {50}},\
  \bibinfo {pages} {91} (\bibinfo {year} {2018})},\ \Eprint
  {http://arxiv.org/abs/1805.11356} {arXiv:1805.11356 [gr-qc]} \BibitemShut
  {NoStop}%
\bibitem [{\citenamefont {Li}\ \emph {et~al.}(2020{\natexlab{a}})\citenamefont
  {Li}, \citenamefont {Singh},\ and\ \citenamefont {Wang}}]{Li:2019}%
  \BibitemOpen
  \bibfield  {author} {\bibinfo {author} {\bibfnamefont {B.-F.}\ \bibnamefont
  {Li}}, \bibinfo {author} {\bibfnamefont {P.}~\bibnamefont {Singh}}, \ and\
  \bibinfo {author} {\bibfnamefont {A.}~\bibnamefont {Wang}},\ }\href {\doibase
  10.1103/PhysRevD.101.086004} {\bibfield  {journal} {\bibinfo  {journal}
  {Phys. Rev. D}\ }\textbf {\bibinfo {volume} {101}},\ \bibinfo {pages}
  {086004} (\bibinfo {year} {2020}{\natexlab{a}})},\ \Eprint
  {http://arxiv.org/abs/1912.08225} {arXiv:1912.08225 [gr-qc]} \BibitemShut
  {NoStop}%
\bibitem [{\citenamefont {Li}\ \emph {et~al.}(2020{\natexlab{b}})\citenamefont
  {Li}, \citenamefont {Olmedo}, \citenamefont {Singh},\ and\ \citenamefont
  {Wang}}]{Li:2020}%
  \BibitemOpen
  \bibfield  {author} {\bibinfo {author} {\bibfnamefont {B.-F.}\ \bibnamefont
  {Li}}, \bibinfo {author} {\bibfnamefont {J.}~\bibnamefont {Olmedo}}, \bibinfo
  {author} {\bibfnamefont {P.}~\bibnamefont {Singh}}, \ and\ \bibinfo {author}
  {\bibfnamefont {A.}~\bibnamefont {Wang}},\ }\href {\doibase
  10.1103/PhysRevD.102.126025} {\bibfield  {journal} {\bibinfo  {journal}
  {Phys. Rev. D}\ }\textbf {\bibinfo {volume} {102}},\ \bibinfo {pages}
  {126025} (\bibinfo {year} {2020}{\natexlab{b}})},\ \Eprint
  {http://arxiv.org/abs/2008.09135} {arXiv:2008.09135 [gr-qc]} \BibitemShut
  {NoStop}%
\bibitem [{\citenamefont {Olmedo}\ and\ \citenamefont
  {Alesci}(2019)}]{Olmedo:2018}%
  \BibitemOpen
  \bibfield  {author} {\bibinfo {author} {\bibfnamefont {J.}~\bibnamefont
  {Olmedo}}\ and\ \bibinfo {author} {\bibfnamefont {E.}~\bibnamefont
  {Alesci}},\ }\href {\doibase 10.1088/1475-7516/2019/04/030} {\bibfield
  {journal} {\bibinfo  {journal} {JCAP}\ }\textbf {\bibinfo {volume} {04}},\
  \bibinfo {pages} {030} (\bibinfo {year} {2019})},\ \Eprint
  {http://arxiv.org/abs/1811.04327} {arXiv:1811.04327 [gr-qc]} \BibitemShut
  {NoStop}%
\bibitem [{\citenamefont {Agullo}\ \emph
  {et~al.}(2021{\natexlab{a}})\citenamefont {Agullo}, \citenamefont {Kranas},\
  and\ \citenamefont {Sreenath}}]{Agullo:2020}%
  \BibitemOpen
  \bibfield  {author} {\bibinfo {author} {\bibfnamefont {I.}~\bibnamefont
  {Agullo}}, \bibinfo {author} {\bibfnamefont {D.}~\bibnamefont {Kranas}}, \
  and\ \bibinfo {author} {\bibfnamefont {V.}~\bibnamefont {Sreenath}},\ }\href
  {\doibase 10.1088/1361-6382/abc521} {\bibfield  {journal} {\bibinfo
  {journal} {Class. Quant. Grav.}\ }\textbf {\bibinfo {volume} {38}},\ \bibinfo
  {pages} {065010} (\bibinfo {year} {2021}{\natexlab{a}})},\ \Eprint
  {http://arxiv.org/abs/2006.09605} {arXiv:2006.09605 [astro-ph.CO]}
  \BibitemShut {NoStop}%
\bibitem [{\citenamefont {Agullo}\ \emph
  {et~al.}(2021{\natexlab{b}})\citenamefont {Agullo}, \citenamefont {Kranas},\
  and\ \citenamefont {Sreenath}}]{Agullo:2020b}%
  \BibitemOpen
  \bibfield  {author} {\bibinfo {author} {\bibfnamefont {I.}~\bibnamefont
  {Agullo}}, \bibinfo {author} {\bibfnamefont {D.}~\bibnamefont {Kranas}}, \
  and\ \bibinfo {author} {\bibfnamefont {V.}~\bibnamefont {Sreenath}},\ }\href
  {\doibase 10.1007/s10714-020-02778-9} {\bibfield  {journal} {\bibinfo
  {journal} {Gen. Rel. Grav.}\ }\textbf {\bibinfo {volume} {53}},\ \bibinfo
  {pages} {17} (\bibinfo {year} {2021}{\natexlab{b}})},\ \Eprint
  {http://arxiv.org/abs/2005.01796} {arXiv:2005.01796 [astro-ph.CO]}
  \BibitemShut {NoStop}%
\bibitem [{\citenamefont {Ashtekar}\ \emph {et~al.}(2020)\citenamefont
  {Ashtekar}, \citenamefont {Gupt}, \citenamefont {Jeong},\ and\ \citenamefont
  {Sreenath}}]{Ashtekar:2020}%
  \BibitemOpen
  \bibfield  {author} {\bibinfo {author} {\bibfnamefont {A.}~\bibnamefont
  {Ashtekar}}, \bibinfo {author} {\bibfnamefont {B.}~\bibnamefont {Gupt}},
  \bibinfo {author} {\bibfnamefont {D.}~\bibnamefont {Jeong}}, \ and\ \bibinfo
  {author} {\bibfnamefont {V.}~\bibnamefont {Sreenath}},\ }\href {\doibase
  10.1103/PhysRevLett.125.051302} {\bibfield  {journal} {\bibinfo  {journal}
  {Phys. Rev. Lett.}\ }\textbf {\bibinfo {volume} {125}},\ \bibinfo {pages}
  {051302} (\bibinfo {year} {2020})},\ \Eprint
  {http://arxiv.org/abs/2001.11689} {arXiv:2001.11689 [astro-ph.CO]}
  \BibitemShut {NoStop}%
\bibitem [{\citenamefont {Ashtekar}\ \emph {et~al.}(2021)\citenamefont
  {Ashtekar}, \citenamefont {Gupt},\ and\ \citenamefont
  {Sreenath}}]{Ashtekar:2021}%
  \BibitemOpen
  \bibfield  {author} {\bibinfo {author} {\bibfnamefont {A.}~\bibnamefont
  {Ashtekar}}, \bibinfo {author} {\bibfnamefont {B.}~\bibnamefont {Gupt}}, \
  and\ \bibinfo {author} {\bibfnamefont {V.}~\bibnamefont {Sreenath}},\
  }\href@noop {} {\  (\bibinfo {year} {2021})},\ \Eprint
  {http://arxiv.org/abs/2103.14568} {arXiv:2103.14568 [gr-qc]} \BibitemShut
  {NoStop}%
\bibitem [{\citenamefont {Han}\ \emph {et~al.}(2020)\citenamefont {Han},
  \citenamefont {Li},\ and\ \citenamefont {Liu}}]{Han:2020}%
  \BibitemOpen
  \bibfield  {author} {\bibinfo {author} {\bibfnamefont {M.}~\bibnamefont
  {Han}}, \bibinfo {author} {\bibfnamefont {H.}~\bibnamefont {Li}}, \ and\
  \bibinfo {author} {\bibfnamefont {H.}~\bibnamefont {Liu}},\ }\href {\doibase
  10.1103/PhysRevD.102.124002} {\bibfield  {journal} {\bibinfo  {journal}
  {Phys. Rev. D}\ }\textbf {\bibinfo {volume} {102}},\ \bibinfo {pages}
  {124002} (\bibinfo {year} {2020})},\ \Eprint
  {http://arxiv.org/abs/2005.00883} {arXiv:2005.00883 [gr-qc]} \BibitemShut
  {NoStop}%
\bibitem [{\citenamefont {Schander}\ and\ \citenamefont
  {Thiemann}(2019)}]{Schander4:2019}%
  \BibitemOpen
  \bibfield  {author} {\bibinfo {author} {\bibfnamefont {S.}~\bibnamefont
  {Schander}}\ and\ \bibinfo {author} {\bibfnamefont {T.}~\bibnamefont
  {Thiemann}},\ }\href@noop {} {\  (\bibinfo {year} {2019})},\ \Eprint
  {http://arxiv.org/abs/1909.07271} {arXiv:1909.07271 [gr-qc]} \BibitemShut
  {NoStop}%
\end{thebibliography}%

\end{document}